\documentclass[journal]{IEEEtran}

\usepackage{cite,graphicx,amsmath,amssymb}
\usepackage{subfigure}
\usepackage{citesort}
\usepackage{fancyhdr}
\usepackage{mdwmath}
\usepackage{mdwtab}
\usepackage{balance}
\usepackage{xcolor}
\usepackage{bm}
\usepackage{amssymb}
\usepackage{amsmath}
\usepackage{cite}
\usepackage{url}
\usepackage{xcolor}
\usepackage{cite,graphicx,amsmath,amssymb}
\usepackage{subfigure}
\usepackage{citesort}
\usepackage{fancyhdr}
\usepackage{mdwmath}
\usepackage{mdwtab}
\usepackage{caption}
\usepackage{amsthm}
\usepackage{algorithm}
\usepackage{algorithmic}
\usepackage{multirow}
\usepackage{flafter}
\usepackage{graphicx}
\usepackage{subfigure}
\usepackage{algorithm}
\usepackage{algorithmic}

\newtheorem{remark}{Remark}

\usepackage{amsmath,amsthm,amssymb}

\newtheorem{theorem}{Theorem}
\newtheorem{lemma}{Lemma}

\makeatletter
\def\ScaleIfNeeded{%
\ifdim\Gin@nat@width>\linewidth \linewidth \else \Gin@nat@width
\fi } \makeatother

\begin{document}

\title{Trajectory Design and Power Control for Multi-UAV Assisted Wireless Networks: A Machine Learning Approach}

\author{
Xiao~Liu,~\IEEEmembership{Student Member,~IEEE,}
 Yuanwei~Liu,~\IEEEmembership{Senior Member,~IEEE,}\\
        Yue~Chen,~\IEEEmembership{Senior Member,~IEEE,}
       and Lajos Hanzo ,~\IEEEmembership{Fellow,~IEEE,}

\thanks{

X. Liu, Y. Liu, and Y. Chen are with the School of Electronic Engineering and Computer Science, Queen Mary University of London, London E1 4NS, UK. (email: x.liu@qmul.ac.uk; yuanwei.liu@qmul.ac.uk; yue.chen@qmul.ac.uk).

L. Hanzo is with the School of Electronics and Computer Science, University of Southampton, SO17 1BJ, UK (e-mail:lh@ecs.soton.ac.uk).

Part of this work has been submitted to IEEE Global Communication Conference (Globecom) 2019~\cite{xiao2019Globecom}.
}
}

 \maketitle

\vspace{-1cm}

\begin{abstract}
A novel framework is proposed for the trajectory design of multiple unmanned aerial vehicles (UAVs) based on the prediction of users' mobility information. The problem of joint trajectory design and power control is formulated for maximizing the instantaneous sum transmit rate while satisfying the rate requirement of users. In an effort to solve this pertinent problem, a three-step approach is proposed which is based on machine learning techniques to obtain both the position information of users and the trajectory design of UAVs. Firstly, a multi-agent Q-learning based placement algorithm is proposed for determining the optimal positions of the UAVs based on the initial location of the users. Secondly, in an effort to determine the mobility information of users based on a real dataset, their position data is collected from Twitter to describe the anonymous user-trajectories in the physical world. In the meantime, an echo state network (ESN) based prediction algorithm is proposed for predicting the future positions of users based on the real dataset. Thirdly, a multi-agent Q-learning based algorithm is conceived for predicting the position of UAVs in each time slot based on the movement of users. In this algorithm, multiple UAVs act as agents to find optimal actions by interacting with their environment and learn from their mistakes. Additionally, we also prove that the proposed multi-agent Q-learning based trajectory design and power control algorithm can converge under mild conditions. Numerical results are provided to demonstrate that as the size of the reservoir increases, the proposed ESN approach improves the prediction accuracy. Finally, we demonstrate that throughput gains of about $17\%$ are achieved.
\end{abstract}


\section{Introduction}
\subsection{Motivation}
As a benefit of their agility, as well as line-of-sight (LoS) propagation, \textit {unmanned aerial vehicles (UAVs)} have  received significant research interests as a means of mitigating a wide range of challenges in commercial and civilian applications~\cite{Zhou2018TVT,khawaja2017uav}. The future wireless communication systems are expected to meet unprecedented demands for high quality wireless services, which imposes challenges on the conventional terrestrial communication networks, especially in traffic hotspots such as in a football stadium or rock concert~\cite{cheng2018TVT,osseiran2014COM,zeng2016COM}. UAVs may be relied upon as aerial base stations to complement and/or support the existing terrestrial communication infrastructure~\cite{wang2018joint,osseiran2014COM,yi2019unified} since they can be flexibly redeployed in temporary traffic hotspots or after natural disasters. Secondly, UAVs have also been deployed as relays between ground-based terminals and as aerial base stations for enhancing the link performance~\cite{kandeepan2014aerial}. Thirdly, UAVs can also be used as aerial base stations to collect data from Internet of Things (IoT) devices on the ground, where building a complete cellular infrastructure is unaffordable~\cite{wang2018joint,Yang2018energy}. Fourthly, combined terrestrial anad UAV communication networks are capable of substantially improving the reliability, security, coverage and throughput of the existing point-to-point UAV-to-ground communications~\cite{zhang2018cellular}.

Key examples of recent advance include the Google Loon project~\cite{katikala2014google}, Facebook's Internet-delivery drone~\cite{Mozaffari2017IEEE_J_WCOM}, and the  AT\&T project of~\cite{zhang2017Optimal}. The drone manufacturing industry faces both opportunities and challenges in the design of UAV-assisted wireless networks. Before fully reaping all the aforementioned benefits, several technical challenges have to be tackled, including the optimal three dimensional (3D) deployment of UAVs, their interference management~\cite{mozaffari2015drone,van2016lte}, energy supply~\cite{zeng2017WCOM,kandeepan2014aerial}, trajectory design~\cite{liu2018comp}, the channel model between the UAV and users~\cite{sun2017air,bing2017study}, resource allocation~\cite{wang2018joint}, as well as the compatibility with the existing infrastructure.

The wide use of online social networks over smartphones has accumulated a rich set of geographical data that describes the anonymous users' mobility information in the physical world~\cite{Zhang2018Synergy}. Many social networking applications like Facebook, Twitter, Wechat, Weibo, etc allow users to 'check-in' and explicitly share their locations, while some other applications have implicitly recorded the users' GPS coordinates~\cite{Yang2016Estimating}, which holds the promise of estimating the geographic user distribution for improving the performance of the system. Reinforcement learning has seen increasing applications in next-generation wireless networks~\cite{Galindo2010Distributed}. More expectantly, reinforcement learning models may be trained by interacting with an environment (states), and they can be expected to find the optimal behaviors (actions) of agents by exploring the environment in an iterative manner and by learning from their mistakes. The model is capable of monitoring the reward resulting form its actions and is chosen for solving problems in UAV-assisted wireless networks.

\subsection{Related Works}
\vspace{-0.05cm}
\subsubsection{Deployment of UAVs}
Among all these challenges, the geographic UAV deployment problems are fundamental. Early research contributions have studied the deployment of a single UAV either to provide maximum radio coverage on the ground~\cite{kosmerl2014ICC,alhourani2014WCOML} or to maximize the number of users by using the minimum transmit power~\cite{alzenad2017WCOML}. As the research evolves further, UAV-assisted systems have received significant attention and been combined with other promising technologies. Specifically, the authors of~\cite{Sharma2017UAV,Sohail2018JA,hou2018multiple} employed non-orthogonal multiple access (NOMA) for improving the performance of UAV-enabled communication systems, which is capable of outperforming orthogonal multiple access (OMA). In~\cite{Mozaffari2016WCOM}, UAV-aided D2D communications was investigated and the tradeoff between the coverage area and the time required for covering the entire target area (delay) by UAV-aided data acquisition was also analyzed. The authors of~\cite{Mozaffari2017IEEE_J_WCOM} proposed a framework using multiple static UAVs for maximizing the average data rate provided for users, while considering fairness amongst the users. The authors of~\cite{mozaffari2016COML} used sphere packing theory for determining the most appropriate 3D position of the UAVs while jointly maximizing both the total coverage area and the battery operating period of the UAVs.

\subsubsection{Trajectory design of UAVs}

It is intuitive that moving UAVs are capable of improving the coverage provided by static UAVs, yet the existing research has mainly considered the scenario that users are static~\cite{Lyu2016Cyclical,Yang2018energy}. Having said that, authors of~\cite{Wu2018trajectory} jointly considered the UAV trajectory and transmit power optimization problem for maintaining fairness among users. An iterative algorithm was invoked for solving the resultant non-convex problem by applying the classic block coordinate descent and successive convex optimization techniques. In~\cite{zeng2017WCOM}, the new design paradigm of jointly optimizing the communication throughput and the UAV's energy consumption was conceived for the determining trajectory of UAV, including its initial/final locations and velocities, as well as its minimum/maximum speed and acceleration. In~\cite{Yang2018energy}, a pair of practical UAV trajectories, namely the circular flight and straight flight were pursued for collecting a given amount of data from a ground terminal (GT) at a fixed location, while considering the associated energy dissipation tradeoff. By contrast, a novel cyclical trajectory was considered in~\cite{Lyu2016Cyclical} to serve each user via TDMA. As shown in~\cite{Lyu2016Cyclical}, a significant throughput gain was achieved over a static UAV. In~\cite{wu2018common}, a simple circular trajectory was used along with maximizing the minimum average throughput of all users. In addition to designing the UAV's trajectory for its action as an aerial base station, the authors of~\cite{zhang2017cellular} studied a cellular-enabled UAV communication system, in which the UAV flew from an initial location to a final location, while maintaining reliable wireless connection with the cellular network by associating the UAV with one of the ground base stations (GBSs) at each time instant. The design-objective was to minimize the UAV's mission completion time by optimizing its trajectory.

\subsection{Our New Contributions}
The aforementioned research contributions considered the deployment and trajectory design of UAVs in the scenario that users are static or studied the movement of UAVs based on the current user location information, where only the user location information of the current time slot is known. Studying the pre-deployment of UAVs based on the full user location information implicitly assumes that the position and mobility information of users is known or it can be predicted. With this proviso the flight trajectory of UAVs may be designed in advance for maintaining a high service quality and hence reduce the response time. Meanwhile, no interaction is needed between the UAVs and ground control center after the pre-deployment of UAVs. To the best of our knowledge, this important problem is still unsolved.

Again, deploying UAVs as aerial BSs is able to provide reliable services for the users~\cite{zhang2017cellular}. However, there is a paucity of research on the problem of 3D trajectory design of multiple UAVs based on the prediction of the users' mobility information, which motivates this treatise. More particularly, i) most existing research contributions mainly focus on the 2D placement of multiple UAVs or on the movement of a single UAV in the scenario, where the users are static. ii) the prediction of the users' position and their mobility information based on a real dataset has never been considered, which helps us to design the trajectory of UAVs in advance, thus reducing both the response time and the interaction between the UAVs as well as control center. the transmit power of UAVs is controlled for obtaining a tradeoff between the received signal power and the interference power, which in turn increases the received signal-interference-noise-rate (SINR). Therefore, we formulate the problem of joint trajectory design and power control of UAVs to improve the users' throughput, while satisfying the rate requirement of users. Against the above background, the primary contributions of this paper are as follows:

\begin{itemize}
\item We propose a novel framework for the trajectory design of multiple UAVs, in which the UAVs move around in a 3D space to offer down-link service to users. Based on the proposed model, we formulate on throughput maximization problem by designing the trajectory and power control of multiple UAVs.
\item We develop a three-step approach for solving the proposed problem. More particularly, i) we propose a multi-agent Q-learning based placement algorithm for determining the initial deployment of UAVs; ii) we propose an echo state network based prediction algorithm for predicting the mobility of users; iii) we conceive a multi-agent Q-learning based trajectory-acquisition and power-control algorithm for UAVs.
\item We invoke the ESN algorithm for acquiring the mobility information of users relying on a real dataset of users collected from Twitter, which consists of  the GPS coordinates and recorded time stamps of Twitter.
\item We conceive a multi-agent Q-learning based solution for the joint trajectory design and power control problem of UAVs. In contrast to a single-agent Q-learning algorithm, the multi-agent Q-learning algorithm is capable of supporting the deployment of cooperative UAVs. We also demonstrate that the proposed algorithms is capable of converging to an optimal state.
\end{itemize}

\subsection{Organization and Notations}

The rest of the paper is organized as follows. In Section II, the problem formulation of joint trajectory design and power control of UAVs is presented. In Section III, the prediction of the users' mobility information is proposed, relying on the ESN algorithm. In Section IV, our multi-agent Q-learning based deployment algorithm is proposed for designing the trajectory and power control of UAVs. Our numerical results are presented in Section V, which is followed by our conclusions in Section VI. The list of notations is illustrated in Table~\ref{List of Notations}.

\begin{table*}[htbp]
\caption{List of Notations}
\centering
\begin{tabular}{|l||r|l||r|}
\hline
\centering
 Notations & Description & Notations & Description\\
\hline
\centering
 $N_u$ & Number of users & $N$ & Number of clusters and UAVs\\
\hline
\centering
 ${x_{{k_n}}},{y_{{k_n}}}$ & Coordinate of users & ${x_n},{y_n}$ & Coordinate of UAVs\\
\hline
\centering
 $f_c$ & Carrier frequency & $h_n$ & Altitude of UAVs\\
\hline
\centering
 ${P_{\max }}$ & UAV transmit power & ${g_{{k_n}}}$ & channel power gain\\
\hline
\centering
 $N_0$ & Noise power spectral & $B$ & Bandwidth \\
\hline
\centering
 ${\mu _{LoS}},{\mu _{NLoS}}$ & Additional path loss for LoS and NLoS & ${P_{LoS}},{P_{NLoS}}$ & LoS and NLoS probability\\
\hline
\centering
 $r_0$ & Minimum rate requirement & ${I_{{k_n}}}$ & Receives interference of users\\
\hline
\centering
 ${r_{{k_n}}}$ & Instantaneous achievable rate & ${R_{sum}}$ & Overall achievable sum rate\\
\hline
\centering
 $b_1$,$b_2$ & Environmental parameters (dense urban) & $\alpha $ & Path loss exponent\\
\hline
\centering
 $N_x$ & Size of neuron reservoir & $a_t$ & State in Q-learning algorithm\\
 \hline
\centering
 $a_t$ & Action in Q-learning algorithm & $r_t$ & Reward in Q-learning algorithm\\
 \hline
\end{tabular}
\label{List of Notations}
\end{table*}

\section{System Model}

We consider the downlink of UAV-assisted wireless communication networks. Multiple UAVs are deployed as aerial BSs to support the users in a particular area, where the terrestrial infrastructure was destroyed or had not been installed. The users are partitioned into $N$ clusters and each user belongs to a single cluster.
Users in this particular area are denoted as $K = \left\{ {{K_1}, \cdots {K_N}} \right\}$, where ${K_n}$ is the set of users that belong to the $n$-th cluster, $n \in \mathbb{N} = \left\{ {1,2, \cdots N} \right\}$. Then, we have ${K_n} \cap {K_{{n'}}} = \phi ,{n'} \ne n,\forall {n'},n \in \mathbb{N}$, while ${{\rm K}_n} = \left| {{K_n}} \right|$ denotes the number of users in the $n$-th cluster. For any cluster $n$, $n \in \mathbb{N}$, we consider a UAV-enabled FDMA system~\cite{he2018joint}, where the UAVs are connected to the core network by satellite. At any time during the UAVs' working period of $T_n$, each UAV communicates simultaneously with multiple users by employing FDMA.

We assume that the energy of UAVs is supplied by laser charging as detailed in~\cite{Liu2016Charging}. A compact distributed laser charging (DLC) receiver can be mounted on a battery-powered off-the-shelf UAV for charging the UAV's battery. A DLC transmitter (termed as a power base station) on the ground is assumed to provide a laser based power supply for the UAVs. Since the DLC is capable of self-alignment and a LOS propagation is usually available because of the high altitude of UAVs, the UAVs can be charged as long as they are flying within the DLC's coverage range. Thus, these DLC-equipped UAVs can operate for a long time without landing until maintenance is needed. The scenario that the energy of UAVs is limited will be discussed in our future work, in which DLC will also be utilized.

\begin{figure} [t!]
\centering
\includegraphics[width=3.2in]{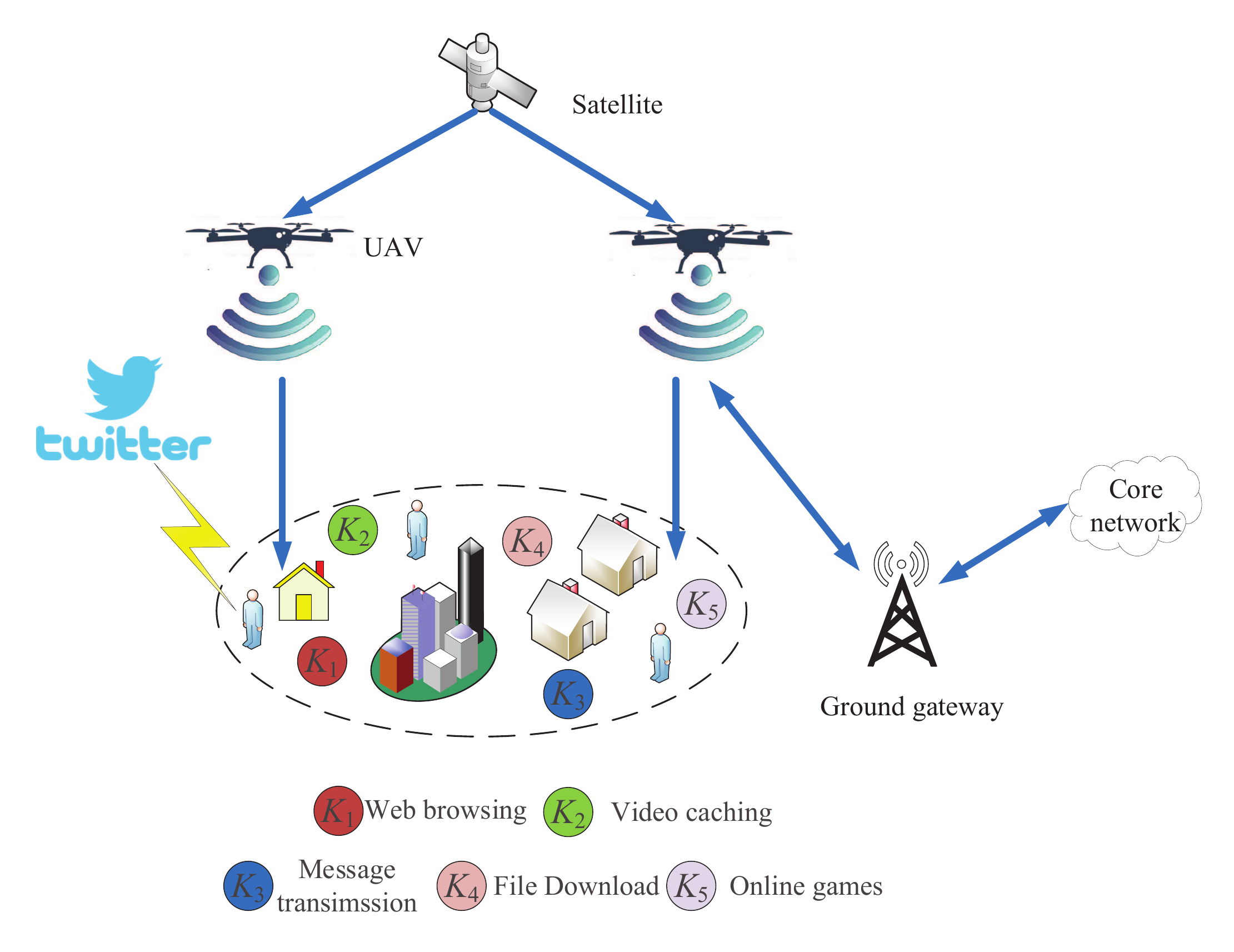}
 \caption{Deployment of multiple UAVs in wireless communications based on the mobility information of users.}\label{qos}
\end{figure}

\subsection{Mobility Model}
Since the users are able to move continuously during the flying period of UAVs, the UAVs have to travel based on the tele-traffic of users. Datasets can be collected to model the mobility of users. Again, in this work, the real-time position information of users is collected from Twitter by the Twitter API, where the data consists of the GPS coordinates and recorded time stamps. When users post tweets, their GPS coordinates are recorded, provided that they give their consent, for example in exchange for calling credits. The detailed discussion of the data collection process is in Section III. The mobility pattern of each user will then be used to determine the optimal location of each UAV, which will naturally impact the service quality of users. The coordinate of each user can be expressed as ${w_{{k_n}}} = {[{x_{{k_n}}}(t),{y_{{k_n}}}(t)]^T} \in {\mathbb{R}^{2 \times 1}},{k_n} \in {K_n}$, where ${\mathbb{R}^{M \times 1}}$ denotes the $M$-dimensional real-valued vector space, while ${x_{{k_n}}}(t)$ and ${y_{{k_n}}}(t)$ are the X-coordinate and Y-coordinate of user $k_n$ at time $t$, respectively.

Since the users are moving continuously, the location of the UAVs must be adjusted accordingly so as to efficiently serve them. The aim of the model is to design the trajectory of UAVs in advance according to the prediction of the users' movement. At any time slot during the UAVs' flight period, both the vertical trajectory (altitude) and the horizontal trajectory of the UAV can be adjusted to offer a high quality of service. The vertical trajectory is denoted by ${h_n}(t) \in [{h_{\min }},{h_{\max }}],0 \le t \le {T_n}$, while the horizontal one by ${q_n}(t) = {[{x_n}(t),{y_n}(t)]^T} \in {\mathbb{R}^{2 \times 1}}$,with $0 \le t \le {T_n}$. The UAVs' operating period is discretized into $N_T$ equal-length time slots.
\subsection{Transmission Model}
In our model, the downlink between the UAVs and users can be regarded as air-to-ground communications. The LoS condition and Non-Line-of-Sight (NLoS) condition are assumed to be encountered randomly. The LoS probability can be expressed as~\cite{Mozaffari2017IEEE_J_WCOM}
\begin{align}\label{plos}
{{P_{{\text{LoS}}}}({\theta _{{k_n}}}) = {b_1}{(\frac{{180}}{\pi }{\theta _{{k_n}}} - \zeta )^{{b_2}}}},
\end{align}
where ${\theta _{{k_n}}}(t) = {\sin ^{ - 1}}(\frac{{{h_n}(t)}}{{{d_{{k_n}}}(t)}})$ is the elevation angle between the UAV and the user ${k_n}$. Furthermore, $b_1$ and $b_2$ are constant values reflecting the environmental impact, while $\zeta $ is also a constant value which is determined both by the antenna and the environment. Naturally, the NLoS probability is given by ${P_{{\text{NLoS}}}} = 1 - {P_{{\text{LoS}}}}$.

Following the free-space path loss model, the channel's power gain between the UAV and user $k_n$ at instant time $t$ is given by
\begin{align}\label{gt}
{{g_{{k_n}}}(t) = {K_0}^{ - 1}d_{{k_n}}^{ - \alpha }(t){[{P_{{\text{LoS}}}}{\mu _{{\text{LoS}}}} + {P_{{\text{NLoS}}}}{\mu _{{\text{NLoS}}}}]^{ - 1}}},
\end{align}
where ${K_0} = {\left( {\frac{{4\pi {f_c}}}{c}} \right)^2}$, $\alpha $ is the path loss exponent,  ${\mu _{LoS}}$ and ${\mu _{NLoS}}$ are the attenuation factors of the LoS and NLoS links, $f_c$ is the carrier frequency, and finally $c$ is the speed of light.

The distance from UAV $n$ to user $k_n$ at time $t$ is assumed to be a constant that can be expressed as
\begin{align}\label{dt}
{{d_{{k_n}}}(t) = \sqrt {{h_n}^2(t) + {{\left[ {{x_n}(t) - {x_{{k_n}}}(t)} \right]}^2} + {{\left[ {{y_n}(t) - {y_{{k_n}}}(t)} \right]}^2}} }.
\end{align}

The transmit power of UAV $n$ has to obey
\begin{align}\label{power}
{0 \le {P_n}(t) \le {P_{\max }}},
\end{align}
where ${P_{\max }}$ is the maximum allowed transmit power of the UAV. Then the transmit power allocated to user $k_n$ at time $t$ is ${p_{{k_n}}}(t) = {{{P_n}(t)} \mathord{\left/
 {\vphantom {{{P_n}(t)} {\left| {{K_n}} \right|}}} \right.
 \kern-\nulldelimiterspace} {\left| {{K_n}} \right|}}$.

\begin{lemma}\label{transmit power}
In order to ensure that every user is capable of connecting to the UAV-assisted network, the lower bound for the transmit power of UAVs has to satisfy
\begin{align}\label{pmax}
{\begin{gathered}
 {P_{\max }}{\kern 1pt} {\kern 1pt} {\kern 1pt}  \ge \left| {{K_n}} \right|{\mu _{{\text{NLoS}}}}{\sigma ^2}{K_0}\left( {{2^{\left| {{K_n}} \right|{{{r_0}} \mathord{\left/
 {\vphantom {{{r_0}} B}} \right.
 \kern-\nulldelimiterspace} B}}} - 1} \right) \\
\cdot \max \left\{ {{h_1},{h_2}, \cdots {h_n}} \right\}
\end{gathered}}.
\end{align}

\begin{proof}
See Appendix A~.
\end{proof}
\end{lemma}

\textbf{Lemma~\ref{transmit power}} sets out the lower bound of the UAV's transmit power for each users' rate requirement to be satisfied.

\begin{remark}\label{remark:power}
Since the users tend to roam continuously, the optimal position of UAVs is changed during each time slot. In this case, the UAVs may also move to offer a better service. When a particular user supported by UAV A moves closer to UAV B while leaving UAV A, the interference may be increased, hence reducing the received SINR, which emphasizes the importance of accurate power control.
\end{remark}

Accordingly, the received SINR ${\Gamma _{{k_n}}}(t)$ of user $k_n$ connected to UAV $i$ at time $t$ can be expressed as
 \begin{align}\label{sinr}
{{\Gamma _{{k_n}}}(t) = \frac{{{p_{{k_n}}}(t){g_{{k_n}}}(t)}}{{{I_{{k_n}}} + {\sigma ^2}}}},
\end{align}
where ${\sigma ^2} = {B_{{k_n}}}N_0^{}$ with $N_0$ denoting the power spectral density of the additive white Gaussian noise (AWGN) at the receivers. Furthermore, ${I_{{k_n}}}(t) = \sum\limits_{{n'} \ne n} {{p_{{k_{{n'}}}}}(t){g_{{k_{{n'}}}}}(t)} $ is the interference imposed on user $k_n$ at time $t$ by the UAVs, except for UAV $n$.

Then the instantaneous achievable rate of user $k_n$ at time $t$, denoted by ${r_{{k_n}}}(t)$ and expressed in bps/Hz becomes
\begin{align}\label{rt}
{{r_{{k_n}}}(t) = {B_{{k_n}}}{\log _2}(1 + \frac{{{p_{{k_n}}}(t){g_{{k_n}}}(t)}}{{{I_{{k_n}}}(t) + {\sigma ^2}}})}.
\end{align}

The overall achievable sum rate at time $t$ can be expressed as
 \begin{align}\label{Rt}
{{R_{{\text{sum}}}} = \sum\limits_{n = 1}^N {\sum\limits_{{k_n} = 1}^{{{\rm K}_n}} {{r_{{k_n}}}(t)} }}.
\end{align}

\begin{figure} [t!]
\centering
\includegraphics[width=3in]{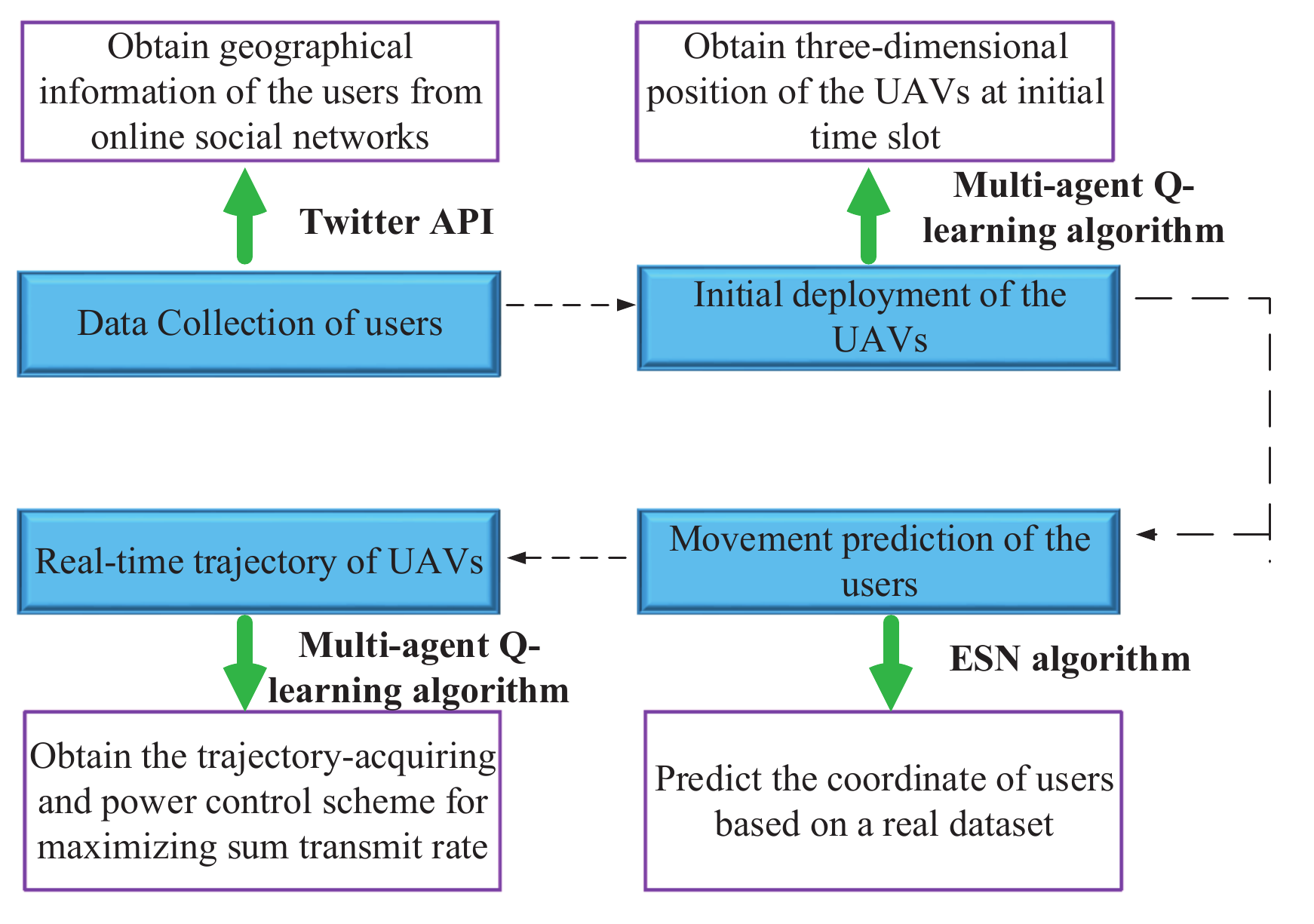}
 \caption{The procedure and algorithms used for solving the joint problem of trajectory plan and power control of UAVs.
 }\label{suanfa}
\end{figure}

\subsection{Problem Formulation}

Let $P = \left\{ {{p_{{k_n}}}(t),{k_n} \in {C_n},0 \le t \le {T_n}} \right\}$, $Q = \left\{ {{q_n}(t),0 \le t \le {T_n}} \right\}$ and $H = \left\{ {{h_n}(t),0 \le t \le {T_n}} \right\}$. Again, we aim for determining both the UAV trajectory and transmit power control at each time slot, i.e., $\left\{ {{P_1}(t),{P_2}(t), \cdots ,{P_n}(t)} \right\}$ and $\left\{ {{x_n}(t),{y_n}(t),{h_n}(t)} \right\},{\kern 1pt} {\kern 1pt} {\kern 1pt} n = 1,2, \cdots N,{\kern 1pt} {\kern 1pt} {\kern 1pt} t = 0,1, \cdots {T_n}$, for maximizing the total transmit rate, while satisfying the rate requirement of each user.

Let us assume that each user's minimum rate requirement $r_0$ is satisfied. This means that all users must have a capacity higher than a rate $r_0$. Our optimization problem is then formulated as
\begin{center}
\begin{subequations}\label{optimizationproblem}
\begin{align}
\mathop {\max }\limits_{C{\text{,}}P{\text{,Q,H}}} {\kern 1pt} {\kern 1pt} {\kern 1pt} {\kern 1pt} {\kern 1pt} {\kern 1pt} {\kern 1pt} {\kern 1pt} {\kern 1pt} {\kern 1pt} {\kern 1pt} {\kern 1pt} {\kern 1pt} {\kern 1pt} {\kern 1pt} {\kern 1pt} {\kern 1pt} {\kern 1pt} {\kern 1pt} {{R_{{\text{sum}}}} = \sum\limits_{n = 1}^N {\sum\limits_{{k_n} = 1}^{{{\rm K}_n}} {{r_{{k_n}}}(t)} }}  \\
{\text{s}}{\text{.t}}{\text{.}}{\kern 1pt} {\kern 1pt} {\kern 1pt} {\kern 1pt}{\kern 1pt} {\kern 1pt} {\kern 1pt} {\kern 1pt} {\kern 1pt} {\kern 1pt} {\kern 1pt} {\kern 1pt} {\kern 1pt} {\kern 1pt}{K_n} \cap {K_{{n'}}} = \phi ,{n'} \ne n,\forall {n'},n \in \mathbb{N},  \\
   {h_{\min }} \le {h_n}(t) \le {h_{\max }},0 \le t \le {T_n}, \\
   {r_{{k_n}}}(t) \ge {r_0},\forall {k_n},t,\hfill \\
   0 \le {P_n}(t) \le {P_{\max }},\forall {k_n},t.
\end{align}
\end{subequations}
\end{center}
where $K(n)$ is the set of users that belong to the cluster $n$, ${h_n}(t)$ is the altitude of UAV $n$ at time slot $t$, while ${P_{{n}}}(t)$ is the total transmit power of UAV $n$ assigned to all users supported by it at time slot $t$. Furthermore, (9b) indicates that each user belongs to a specific cluster which is covered by a single UAV; 9c) formulates the altitude bound of UAVs; (9d) qualifies the rate requirement of each user; (9e) represents the power control constraint of UAVs. Here we note that designing the trajectory of UAVs will ensure that they are in the optimal position at each time slot. This, in turn, will lead to improving the instantaneous transmit rate. Meanwhile, designing the trajectory of UAVs in advance based on the prediction of the users' mobility will also reduce the response time of UAVs, despite reducing the interactions among the UAVs and the ground control center. Fig.2 summarizes the framework proposed for solving the problem considered. Given this framework, we utilize the ESN-based predictions of the users' movement.

\begin{remark}\label{remark:rate}
The instantaneous transmit rate depends on the transmit power, on the number of UAVs, and on the location of UAVs (horizonal position and altitude).
\end{remark}

Problem (9a) is challenging since the objective function is non-convex as a function of ${x_n}(t)$, ${y_n}(t)$ and ${h_n}(t)$~\cite{zeng2017WCOM,wang2018joint}. Indeed it has been shown that problem (9a) is NP-hard even if we only consider the users' clustering~\cite{liu2018globalcom}. Exhaustive search exhibits an excessive complexity. In order to solve this problem at a low complexity, a multi-agent Q-learning algorithm will be invoked in Section IV for finding the optimal solution with a high probability, despite searching through only small fraction of the entire design-space.

\section{Echo State Network Algorithm for Prediction of Users' Movement}
In this section, we formulate our ESN algorithm for predicting the movement of users. A variety of mobility models have been utilized in~\cite{Ren2017Acess,mingzhe2017JSAC}. However, in these mobility models, the direction of each user's movement tends to be uniformly distributed among left, right, forward and backward, which does not fully reflect the real movement of users. In this section, we tackle this problem by predicting the mobility of users based on a real dataset collected from Twitter.
\subsection{Data collection of users}

\begin{figure} [t!]
\centering
\includegraphics[width=3in]{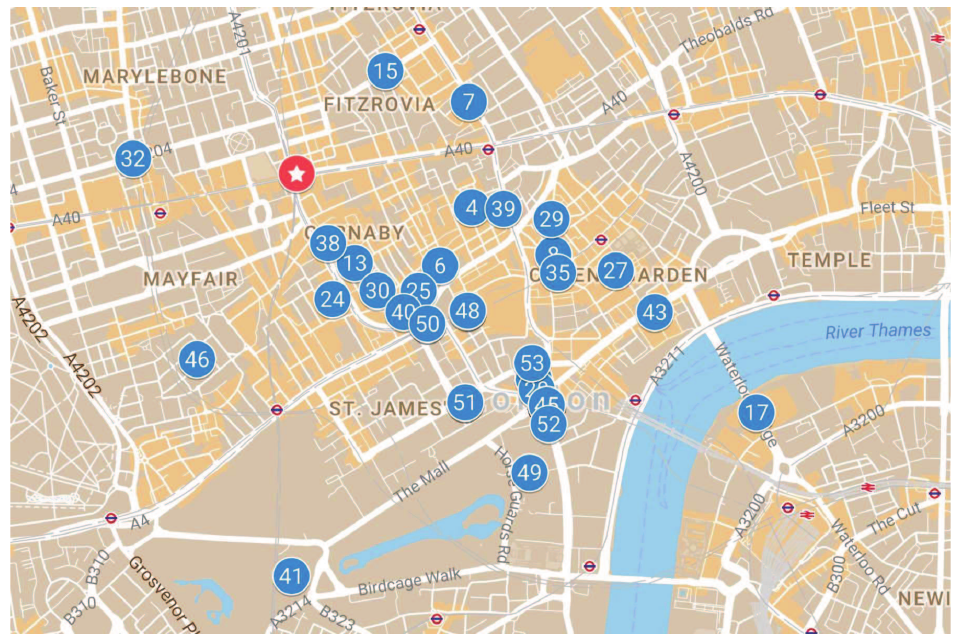}
 \caption{The initial positions of the users derived from Twitter.
 }\label{daymap}
\end{figure}

In order to obtain real mobility information, the relevant position data has to be collected. Serendipitously, the wide use of online social network (OSN) APPs over smartphones has accumulated a rich set of geographical data that describes anonymous user trajectories in the physical world, which holds the promise of providing a lightweight means of studying the mobility of users. For example, many social networking applications like Facebook and Weibo allow users to 'check-in' and explicitly show their locations. Some other applications implicitly record the users' GPS coordinates~\cite{Yang2016Estimating}.

The users' locations can be predicted by mining data from social networks, given that the observed movement is associated with certain reference locations. One of the most effective method of collecting position information relies on the Twitter API. When Twitter users tweet, their GPS-related position information is recorded by the Twitter API and it becomes available to the general public. We relied on 12000 twitter collected near Oxford Street, in London on the 14th, March 2018\footnote[1]{The dataset has been shared by authors in Github. It is shown on the websit: https://github.com/pswf/Twitter-Dataset/blob/master/Dataset. Our approach can accommodate other datasets without loss of generality.}. Among these twitter users, 50 users who tweeted more than 3 times were encountered. In this case, the movement of these 50 users is recorded. Fig.~\ref{daymap} illustrates the distribution of these 50 users at the initial time of collecting data. In an effort to obtain more information about a user to characterise the movement more specifically, classic interpolation methods was used to make sure that the position information of each users were recorded every 200 seconds. In this case, the trajectory of each user during this period was obtained. The position of users during the $n$th time slot can be expressed as $u(n) = {\left[ {{u_1}(n),{u_2}(n),\cdots {u_{{N_u}}}(n)} \right]^T}$, where $N_u$ is the total number of users.

\subsection{Echo State Network Algorithm for the Prediction of Users' Movement}

The ESN model's input is the position vector of users collected from Twitter, namely $u(n) = {\left[ {{u_1}(n),{u_2}(n), \cdots {u_{{N_u}}}(n)} \right]^T}$, while its output vector is the position information of users predicted by the ESN algorithm, namely $y(n) = {\left[ {{y_1}(n),{y_2}(n), \cdots {y_{{N_u}}}(n)} \right]^T}$. For each different user, the ESN model is initialized before it imports in the new inputs. As illustrated in Fig.4, the ESN model essentially consists of three layers: input layer, neuron reservoir and output layer~\cite{mingzhe2017JSAC}. The $W_{in}$ and $W_{out}$ represent the connections between these three layers, represented as matrices. The $W$ is another matrix that presents the connections between the neurons in neuron reservoir. Every segment is fixed once the whole network is established, except $W_{out}$, which is the only trainable part in the network.

The classic mean square error (MSE) metric is invoked for evaluating the prediction accuracy~\cite{mingzhe2017JSAC}
\begin{align}\label{esn1}
{{\text{MSE}}\left( {y,{y^{{\text{target}}}}} \right) = \frac{1}{{{N_u}}}\sum\limits_{n = 1}^{{N_u}} {\sqrt {\frac{1}{T}\sum\limits_{i = 1}^T {{{\left[ {{y_i}(n) - {y_i}^{{\text{target}}}(n)} \right]}^2}} } } }.
\end{align}
where $y$ and ${y^{{\text{target}}}}$ are the predicted and the real position of the users, respectively.

\begin{remark}\label{remark:aim}
The aim of the ESN algorithm is to train a model with the aid of its input and out put to minimizes the MSE.
\end{remark}

\begin{figure} [t!]
\centering
\includegraphics[width=3in]{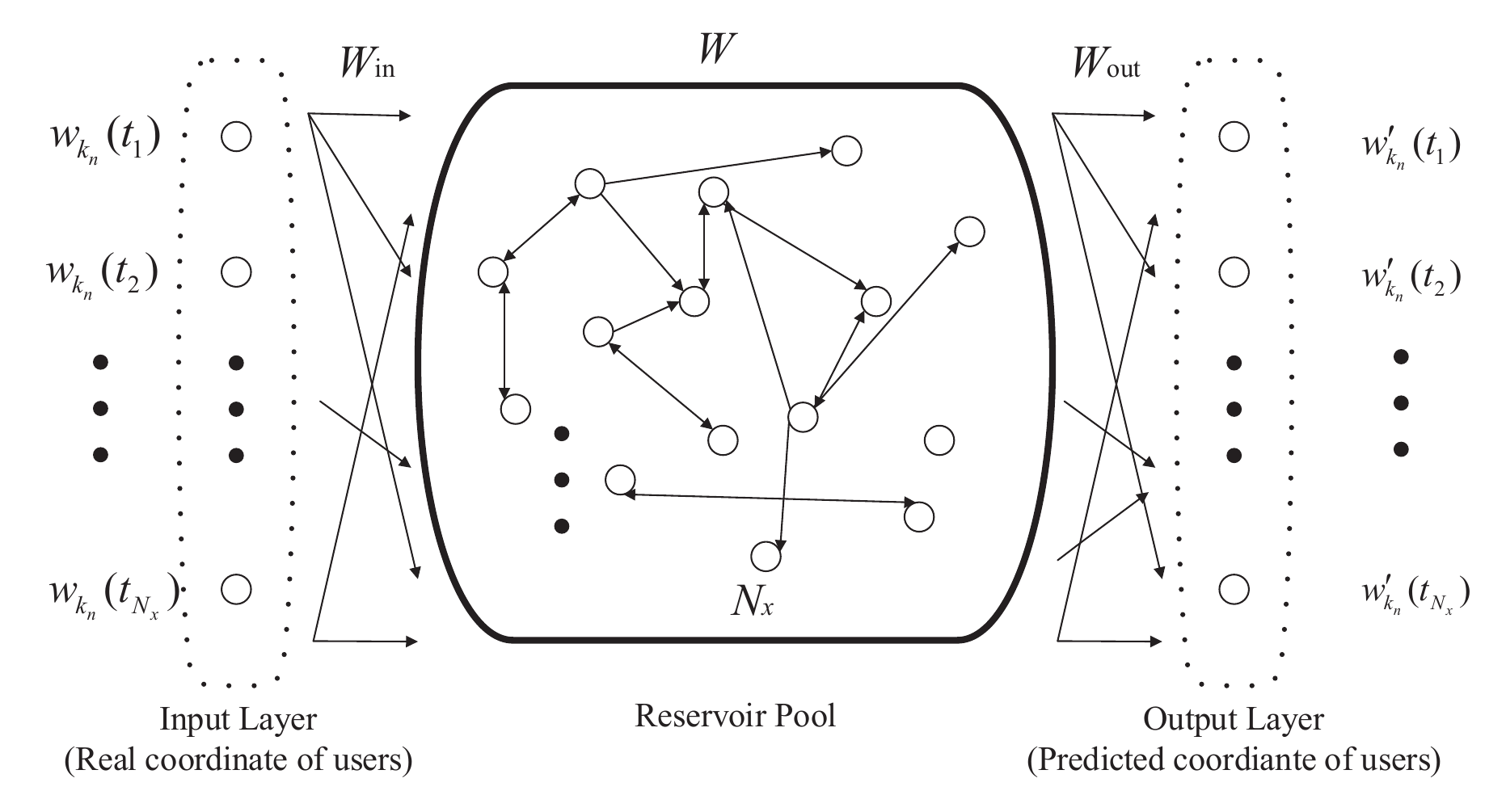}
 \caption{The structure of Echo State Network for predicting the mobility of the users.}\label{esn}
\end{figure}

\begin{algorithm}[!t]
\caption{ESN algorithm for Predicting Users' Movement}
\label{esn}
\begin{algorithmic}[1]
\REQUIRE ~~
75\% of the dataset for training process, 25\% of the dataset for testing process.\\

\STATE  \textbf{Initialize:} $W_j^{a,in},W_j^a,W_j^{a,out},{y_i} = 0$.
\STATE  \textbf{Training stage:}
\STATE  \textbf{for} $i$ from 0 to $N_u$ \textbf{do}
\STATE  \textbf{  }\textbf{  }\textbf{  }\textbf{  }\textbf{for} $n$ from 0 to $N_x$ \textbf{do}
\STATE  \textbf{  }\textbf{  }\textbf{  }\textbf{  }\textbf{  }\textbf{  }\textbf{  }\textbf{  }Computer the update equations according to Eq. (12).
\STATE  \textbf{  }\textbf{  }\textbf{  }\textbf{  }\textbf{  }\textbf{  }\textbf{  }\textbf{  }Update the network outputs according to Eq. (13).
\STATE \textbf{  }\textbf{  }\textbf{  }\textbf{  }\textbf{end for}
\STATE \textbf{end for}
\STATE  \textbf{Prediction stage:}
\STATE  Get the prediction of users' mobility information based on the output weight matrix ${W_{{\text{out}}}}$.
\ENSURE ~~
Predicted coordinate of users.
\end{algorithmic}
\end{algorithm}

The neuron reservoir is a sparse network, which consists of sparsely connected neurons, having a short-term memory of the previous states encountered. In the neuron reservoir, the typical update equations are given by
\begin{align}\label{function2}
\tilde{x}(n) = tanh\Big(W_{in}[0:u(n)] + W\cdot x(n-1)\Big),
\end{align}
\begin{align}\label{function3}
x(n) = (1-\alpha)x(n-1) + \alpha \tilde{x}(n),
\end{align}
where $x(n)\in\mathbb{R}^{N_x}$ is the updated version of the variable $\tilde{x}(n)$, $N_x$ is the size of the neuron reservoir, $\alpha$ is the leakage rate, while $tanh(\cdot)$ is the activation function of neurons in the reservoir. Additionally, $W_{in} \in \mathbb{R}^{N_x \cdot(1+N_u)}$ and $W \in \mathbb{R}^{N_x \cdot N_x}$ are the input and the recurrent weight matrices, respectively. The input matrix $W^{in}$ and the recurrent connection matrices $W$ are randomly generated, while the leakage rate $\alpha$ is from the interval of $[0, 1)$.

After data echoes in the pool, it flows to the output layer, which is characterized as
\begin{align}\label{output function}
y(n) = W_{{\text{out}}}[0;x(n)],
\end{align}
where $y(n)\in\mathbb{R}^{N_y}$ represents the network outputs, while $W_{{\text{out}}}\in\mathbb{R}^{N_y\cdot(1+N_u+N_x)}$ the weight matrix of outputs.

The neuron reservoir is determined by four parameters: the size of the pool, its sparsity, the distribution of its nonzero elements and spectral radius of $W$.

\begin{itemize}
  \item \textbf{Size of Neuron Reservoir $N_x$}: represents the number of neurons in the reservoir, which is the most crucial parameter of the ESN algorithm. The larger $N_x$, the more precise prediction becomes, but at the same time it increases the probability of causing overfitting.
  \item \textbf{Sparsity}: Sparsity characterizes the density of the connections between neurons in the reservoir. When the density is reduced, the non-linear closing capability is increased, whilst the operation becomes more complex.
  \item \textbf{Distribution of Nonzero Elements}: The matrix $W$ is typically a sparse one, representing a network, which has normally distributed elements centered around zero. In this paper, we use a continuous-valued bounded uniform distribution, which provides an excellent performance~\cite{mingzhe2017esn}, outperforming many other distributions.
  \item \textbf{Spectral Radius of \boldmath$W$}: Spectral Radius of \boldmath$W$ scales the matrix $W$ and hence also the variance of its nonzero elements. This parameter is fixed, once the neuron reservoir is established.
\end{itemize}

\begin{remark}\label{remark:capacity}
The size of neuron reservoir has to be carefully chosen to satisfy the memory constraint, but $N_x$ should also be at least equal to the estimate of independent real values the reservoir has to remember from the input in order to solve the task.
\end{remark}

A larger memory capacity implies that the ESN model is capable of storing more locations that the users have visited, which tends to improve the prediction accuracy of the users' movements. In the ESN model, typically 75\% of the dataset is used for training and 25\% for the testing process.
\begin{remark}\label{remark:reservoir}
For challenging tasks, as large a neuron reservoir has to be used as one can computationally afford.
\end{remark}

\section{Joint Trajectory Design and transmit power control of UAVs}

In this section, we assume that in any cluster $n$, the UAV is serving the users relying on an adaptively controlled flight trajectory and transmit power. With the goal of maximizing the sum transmit rate in each time slot by determining the flight trajectory and transmit power of the UAVs. User clustering constitutes the first step of achieving the association between the UAVs and the users. The users are partitioned into different clusters, and each cluster is served by a single UAV. The process of cell partitioning has been discussed in our previous work~\cite{liu2018globalcom}, which has demonstrated that the genetic K-means (GAK-means) algorithm is capable of obtaining globally optimal clustering results. The process of clustering is also detailed in~\cite{liu2018globalcom}, hence it will not be elaborated on here.

\begin{figure} [t!]
\centering
\includegraphics[width=3in]{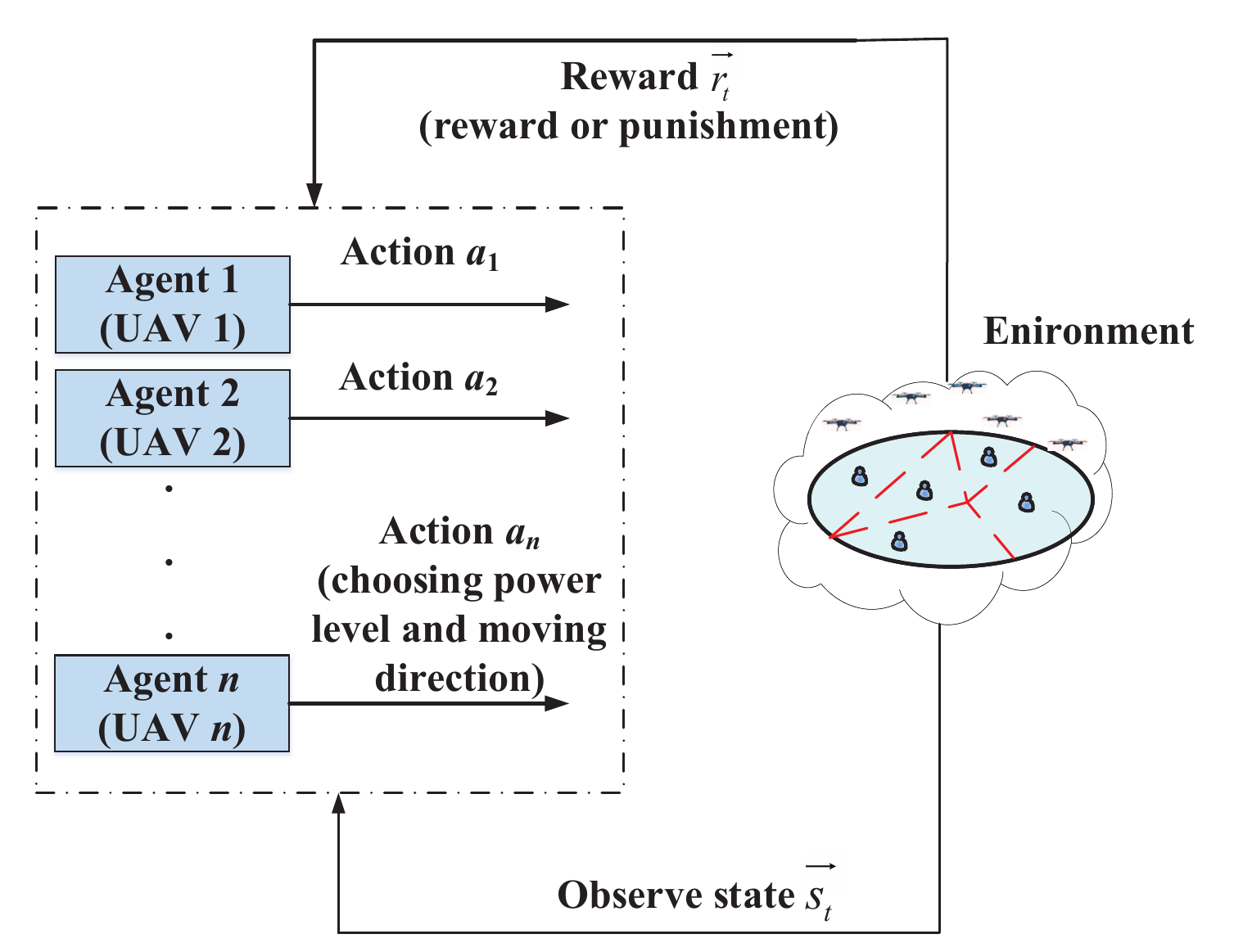}
 \caption{The structure of multi-agent Q-learning for the trajectory design and power control of the UAVs.
 }\label{suanfa}
\end{figure}

\subsection{Signal-Agent Q-learning Algorithm}
In this section, a multi-agent Q-learning algorithm is invoked for obtaining the movement of the UAVs. Before introducing multi-agent Q-learning algorithm, the single agent Q-learning algorithm is introduced as the theoretical basis. In the single agent model, each UAV acts as an agent, moving without cooperating with other UAVs. In this case, the geographic positioning of each UAV is not affected by the movement of other UAVs. The single agent Q-learning model relies on four core elements: the states, actions, rewards and Q-values. The aim of this algorithm is that of conceiving a policy (a set of actions will be carried out by the agent) that maximizes the rewards observed during the interaction time of the agent. During the iterations, the agent observes a state ${s_t}$, in each time slot $t$ from the state space $S$. Accordingly, the agent carries out an action ${a_t}$, from the action space $A$, selecting its specific flying directions and transmit power based on policy $J$. The decision policy $J$ is determined by a Q-table $Q({s_t},{a_t})$. The policy promote choosing specific actions, which enable the model to attain the maximum Q-values. Following each action, the state of the agent traverses to a new state ${s_{t + 1}}$, while the agent receives a reward, ${r_t}$, which is determined by the instantaneous sum rate of users.

\begin{algorithm}[!t]
\caption{The proposed multi-agent Q-learning algorithm for deployment of multiple UAVs}
\label{q1}
\begin{algorithmic}[1]

\STATE  Let $t = 0$, $Q_n^0({s_n},{a_n}) = 0$ for all $s_n$ and $a_n$
\STATE  \textbf{Initialize:} the starting state $s_t$
\STATE  \textbf{Loop:}\\
\STATE  \textbf{  }\textbf{  }\textbf{  }\textbf{  }send $Q_n^t(s_n^t,:)$ to all other cooperating agents $j$
\STATE  \textbf{  }\textbf{  }\textbf{  }\textbf{  }receive $Q_j^t(s_j^t,:)$ from all other cooperating agents $j$
\STATE  \textbf{  }\textbf{  }\textbf{  }\textbf{  }\textbf{if} ${\text{random}} < \varepsilon $ then
\STATE  \textbf{  }\textbf{  }\textbf{  }\textbf{  }\textbf{  }\textbf{  }select action randomly
\STATE  \textbf{  }\textbf{  }\textbf{  }\textbf{  }\textbf{else}
\STATE  \textbf{  }\textbf{  }\textbf{  }\textbf{  }\textbf{  }\textbf{  }\textbf{  }choose action: $a_n^t = \arg {\max _a}\left( {\sum\nolimits_{1 \leqslant j \leqslant N} {Q_j^t\left( {s_j^t,a} \right)} } \right)$
\STATE  \textbf{  }\textbf{  }\textbf{  }\textbf{  }\textbf{  }\textbf{  }receive reward $r_n^t$
\STATE  \textbf{  }\textbf{  }\textbf{  }\textbf{  }\textbf{  }\textbf{  }observe next state $s_n^{t + 1}$
\STATE  \textbf{  }\textbf{  }\textbf{  }\textbf{  }\textbf{  }\textbf{  }update Q-table as ${Q_n}\left( {{s_n},{a_n}} \right) \leftarrow  (1 - \alpha ){Q_n}\left( {{s_n},{a_n}} \right) + \alpha \left( {{r_n}\left( {{s_n},\overrightarrow a } \right) + \beta \mathop {\max {Q_n}\left( {{{s'}_n},b} \right)}\limits_{b \in {A_n}} } \right)$
\STATE  \textbf{  }\textbf{  }\textbf{  }\textbf{  }$s_n^t = s_n^{t + 1}$
\STATE  \textbf{end loop}

\end{algorithmic}
\end{algorithm}

\subsection{State-Action Construction of the Multi-agent Q-learning Algorithm}

In the multi-agent Q-learning model, each agent has to keep a Q-table that includes data both about its own states as well as of the other agents' states and actions. More explicitly, it takes account of the other agents' actions with the goal of promoting cooperative actions among agents so as to glean the highest possible rewards.

In the multi-agent Q-learning model, the individual agents are represented by a four-tuple state: ${\xi _n} = (x_{{\text{UAV}}}^{(n)},y_{{\text{UAV}}}^{(n)},h_{{\text{UAV}}}^{(n)},P_{{\text{UAV}}}^{(n)})$, where $(x_{{\text{UAV}}}^{(n)},y_{{\text{UAV}}}^{(n)})$ is the horizonal position of UAV $n$, while $h_{{\text{UAV}}}^{(n)}$ and $P_{{\text{UAV}}}^{(n)}$ are the altitude and the transmit power of UAV $n$, respectively. Since the UAVs operate across a particular area, the corresponding state space is donated as: $x_{{\text{UAV}}}^{(n)}:\left\{ {0,1, \cdots {X_d}} \right\}$, $y_{{\text{UAV}}}^{(n)}:\left\{ {0,1, \cdots {Y_d}} \right\}$, $h_{{\text{UAV}}}^{(n)}:\left\{ {{h_{\min }}, \cdots {h_{\max }}} \right\}$, $P_{{\text{UAV}}}^{(n)} = \left\{ {0, \cdots {P_{\max }}} \right\}$, where ${X_d}$ and ${Y_d}$ represent the maximum coordinate of this particular area. while $h_{\min }$ and $h_{\max }$ are the lower and upper altitude bound of UAVs, respectively. Finally, $P_{\max }$ is the maximum transmit power derived from \textbf{Lemma 1}.

We assume that the initial state of UAVs is determined randomly. Then the convergence of the algorithm is determined by the number of users and UAVs, as well as by the initial position of UAVs. A faster convergence is attained when the UAVs are placed closer to the respective optimal positions.

   \begin{figure*}
   \begin{align}\label{reward}
{{{r}_{n}}(t)=\left\{ \begin{matrix}
   \sum\limits_{{{k}_{n}}=1}^{{{\text{K}}_{n}}}{{{B}_{{{k}_{n}}}}{{\log }_{2}}\left( 1+\frac{{{p}_{{{k}_{n}}}}(t){{g}_{{{k}_{n}}}}(t)}{{{I}_{{{k}_{n}}}}(t)+{{\sigma }^{2}}} \right),} & \text{Sumrat}{{\text{e}}_{\text{new}}}\ge \text{Sumrat}{{\text{e}}_{\text{old}}},  \\
   0 & \text{Sumrat}{{\text{e}}_{\text{new}}}<\text{Sumrat}{{\text{e}}_{\text{old}}}.  \\
\end{matrix} \right.}
\end{align}
\end{figure*}

At each step, each UAV carries out an action ${a_t} \in A$, which includes choosing a specific direction and transmit power level, depending on its current state, ${s_t} \in S$, based on the decision policy $J$. The UAVs may fly in arbitrary directions (with different angles), which makes the problem non-trivial to solve. However, by assuming the UAVs fly at a constant velocity, and obey coordinated turns, the model may be simplified to as few as 7 directions (left, right, forward, backward, upward, downward and maintaining static). The number of the directions has to be appropriately chosen in practice to strike a tradeoff between the accuracy and algorithmic complexity. Additionally, we assume that the transmit power of the UAVs only has 3 values, namely 0.08W, 0.09W and 0.1W\footnote[2]{In this paper, the proposed algorithm can accommodate any arbitrary number of power level without loss of generality. We choose three power levels to strike a tradeoff between the performance and complexity of the system.}.

\begin{remark}\label{remark:degree}
In the real application of UAVs as aerial base stations, they can fly in arbitrary directions, but we constrain their mobility to as few as 7 directions.
\end{remark}

We choose the 3D position of the UAVs (horizontal coordinates and altitudes) and the transmit power to define their states. The actions of each agent are determined by a set of coordinates for specifying their travel directions and the candidate transmit power of the UAVs\footnote[3]{In our future work, we will consider the online design of UAVs' trajectories, and the mobility of UAVs will be constrained to 360 degree of angles instead of 7 directions. Given that, the state-action space is huge, a deep multi-agent Q-network based algorithm will be proposed in our future work.}. Explicitly, (1,0,0) means that the UAV turns right; (-1,0,0) indicates that the UAV turns left; (0,1,0) represents that the UAV flies forward; (0,-1,0) means that the UAV flies backward; (0,0,1) implies that the UAV rises; (0,0,-1) means that the UAV descends; (0,0,0) indicates that the UAV stays static. In terms of power, we assume 0.08W, 0.09W and 0.1W. Again, we set the initial transmit power to 0.08W, and each UAV carries out an action from the set {increase, decrease and maintain} at each time slot. Then, the entire action space has as few as $3\times 7$ =21 elements.


\subsection{Reward Function of Multi-agent Q-learning Algorithm}

\begin{figure*} \centering
\subfigure[Comparison of coordinate vs time.] { \label{fig:a}
\includegraphics[width=0.88\columnwidth]{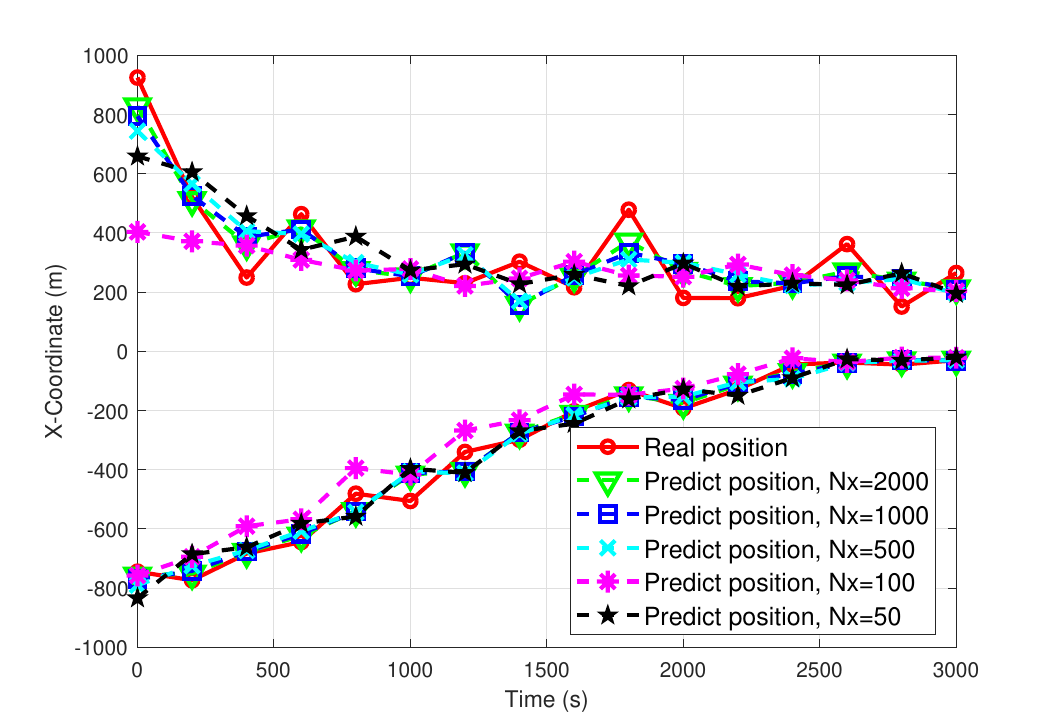}
}
\subfigure[Comparison of tracks.] { \label{fig:b}
\includegraphics[width=0.88\columnwidth]{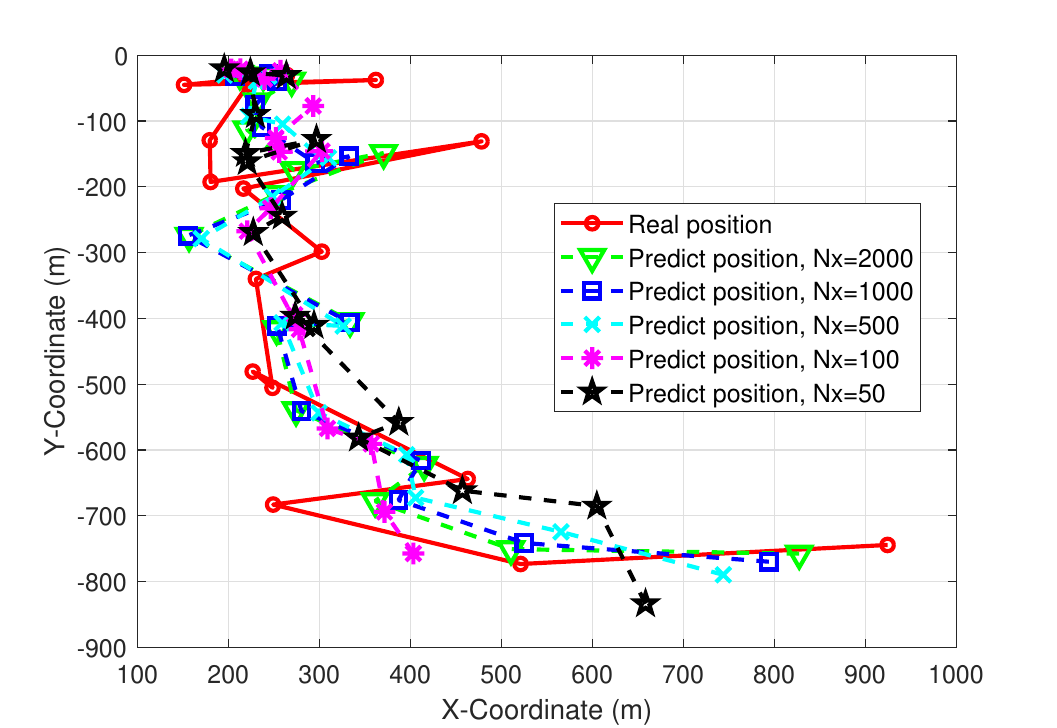}
}
\caption{Comparison of real tracks and predicted tracks for different neuron reservoir size. }\label{esnplot}
\end{figure*}

One of the main limitations of reinforcement learning is its slow convergence. The beneficial design of the reward function requires a sophisticated methodology for accelerating the convergence to the optimal solution~\cite{matignon2006reward}. In the multi-agent Q-learning model, each agent has the same reward or punishment. The reward function is directly related to the instantaneous sum rate of the users. When the UAV carries out an action at time instant $t$, and this action improves the sum rate, then the UAV receives a reward, and vice versa. The global reward function is formulated as \eqref{reward} at the top of next page.

\begin{remark}\label{remark:reward}
Altering the value of reward does not change the final result of the algorithm, but its convergence rate is indeed influenced. Using a continuous reward function is capable of faster convergence than a binary reward function~\cite{matignon2006reward}.
\end{remark}

\subsection{Transition of Multi-agent Q-learning Algorithm}

In this part, we extend the model from single-agent Q-learning to multi-agent Q-learning. First, we redefine the Q-values for the the multi-agent model, and then present the algorithm conceived for learning the Q-values.

To adapt the single-agent model to the multi-agent context, the first step is that of recognizing the joint actions, rather than merely carrying out individual actions. For an $N$-agent system, the Q-function for any individual agent is $Q(s,{a^1}, \cdots {a^N})$, rather than the single-agent Q-function, $Q(s,a)$. Given the extended notion of the Q-function, we define the Q-value as the expected sum of discounted rewards when all agents follow specific strategies from the next period. This definition differs from the single-agent model, where the future rewards are simply based on the agent's own optimal strategy. More precisely, we refer to $Q_*^n$ as the Q-function for agent $n$.

\begin{remark}\label{remark:difference}
The difference of multi-agent model compared to the single-agent model is that the reward function of multi-agent model is dependent on the joint action of all agents $\overrightarrow a $.
\end{remark}

Sparked by \textbf{Remark 8}, the update rule has to obey
\begin{align}\label{mql}
{\begin{gathered}
  {Q_n}\left( {{s_n},{a_n}} \right) \leftarrow  (1 - \alpha ){Q_n}\left( {{s_n},{a_n}} \right) \hfill \\
\textbf{  }\textbf{  }\textbf{  }\textbf{  }\textbf{  }\textbf{  }\textbf{  }\textbf{  }\textbf{  }\textbf{  }\textbf{  }\textbf{  }\textbf{  }\textbf{  }+ \alpha \left( {{r_n}\left( {{s_n},\overrightarrow a } \right) + \beta \mathop {\max {Q_n}\left( {{{s'}_n},b} \right)}\limits_{b \in {A_n}} } \right) \hfill \\
\end{gathered} }.
\end{align}

The $n$th agent shares the row of its Q-table that corresponds to its current state with all other cooperating agents $j$, $j=1,\cdots ,N$. Then the $n$th agent selects its action according to
\begin{align}\label{mqlaction}
{a_n^t = \arg {\max _a}\left( {\sum\nolimits_{1 \le j \le N} {Q_j^t\left( {s_j^t,a} \right)} } \right)}.
\end{align}

In order to carry out multi-agent training, we train one agent at a time, and keep the policies of all the other agents fixed during this period.

The main idea behind this strategy depends on the global Q-value $Q(s,a)$, which represents the Q-value of the whole model. This global Q-value can be decomposed into a linear combination of local agent-dependent Q-values as follows: $Q(s,a) = \sum\nolimits_{1 \le j \le N} {{Q_j}\left( {{s_j},{a_j}} \right)}$. Thus, if each agent $j$ maximizes its own Q-value, the global Q-value will be maximized.

The transition from the current state ${s_t}$ to the state of the next time slot ${s_{t + 1}}$ with reward ${r_t}$ when action ${a_t}$ is taken can be characterized by the conditional transition probability $p({s_{t + 1}},{r_t}|{s_t},{a_t})$. The goal of learning is that of maximizing the gain defined as the expected cumulative discounted rewards
\begin{align}\label{reward}
{{G_t} = {\rm E}[\sum\limits_{n = 0}^\infty  {{\beta ^n}{r_{t + n}}} ]},
\end{align}
where $\beta $ is the discount factor. The model relies on the learning rate $\alpha $, discount factor $\beta $ and a greedy policy $J$ associated with the probability $\varepsilon $ to increase the exploration actions. The learning process is divided into episodes, and the UAVs' state will be re-initialized at the beginning of each episode. At each time slot, each UAV needs to figure out the optimal action for the objective function.

\begin{theorem}\label{convergence}
Multi-agent Q-learning ${\mathbf{MQ}}_{k + 1}^{}$ converges to an optimal state ${\mathbf{M}}{{\mathbf{Q}}^ * }\left[ {k + 1} \right]$, where $k$ is the episode time.
\begin{proof}
See Appendix A~.
\end{proof}
\end{theorem}


\subsection{Complexity of the Algorithm}
The complexity of the algorithm has two main contributors, namely the complexity of the GAK-means based clustering algorithm and that of the multi-agent Q-learning based trajectory-acquisition as well as power control algorithm. In terms of the first one, the proposed scheme involves three steps during each iteration. The first stage calculates the Euclidean distance between each user and cluster centers. For $N_u$ users and $N$ clusters, calculating all Euclidean distances requires on the order of  $O\left( {6K{N_u}} \right)$ floating-point operations. The second stage allocates each user to the specific cluster having the closest center, which requires $O\left[ {{N_u}\left( {N - 1} \right]} \right)$ comparisons.Furthermore, the complexity of recalculating the cluster center is $O\left( {4N{N_u}} \right)$. Therefore, the total computational complexity of GAK-means clustering is on the order of $O\left[ {6N{N_u} + {N_u}\left( {N + 1} \right) + 4N{N_u}} \right] \approx O\left( {N{N_u}} \right)$.

\begin{figure}
\centering
\includegraphics[width=3in]{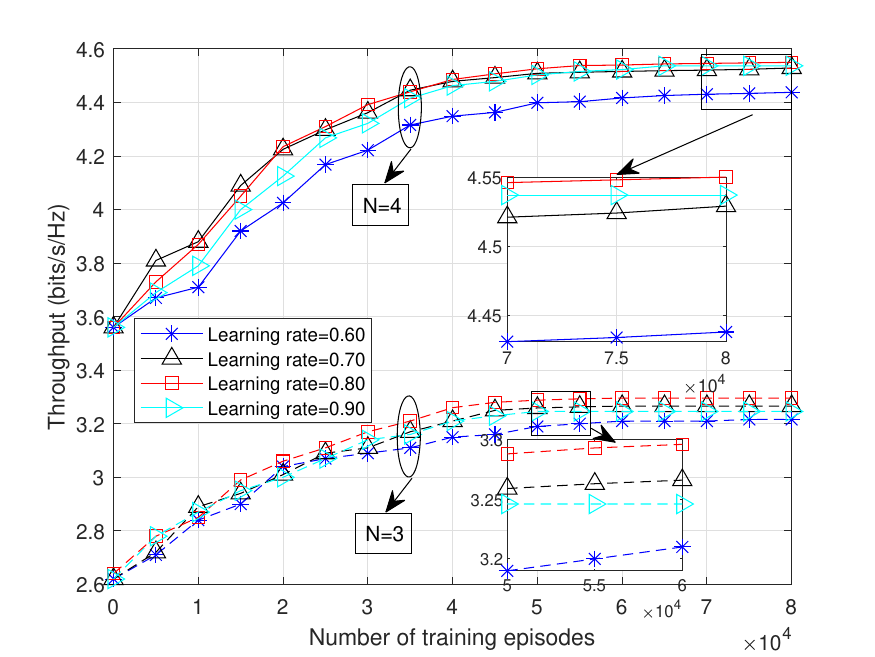}
\caption{Convergence of the proposed algorithm vs the number of training episodes.}\label{convergence2}
\end{figure}

\begin{figure} [t!]
\centering
\includegraphics[width=3.0in]{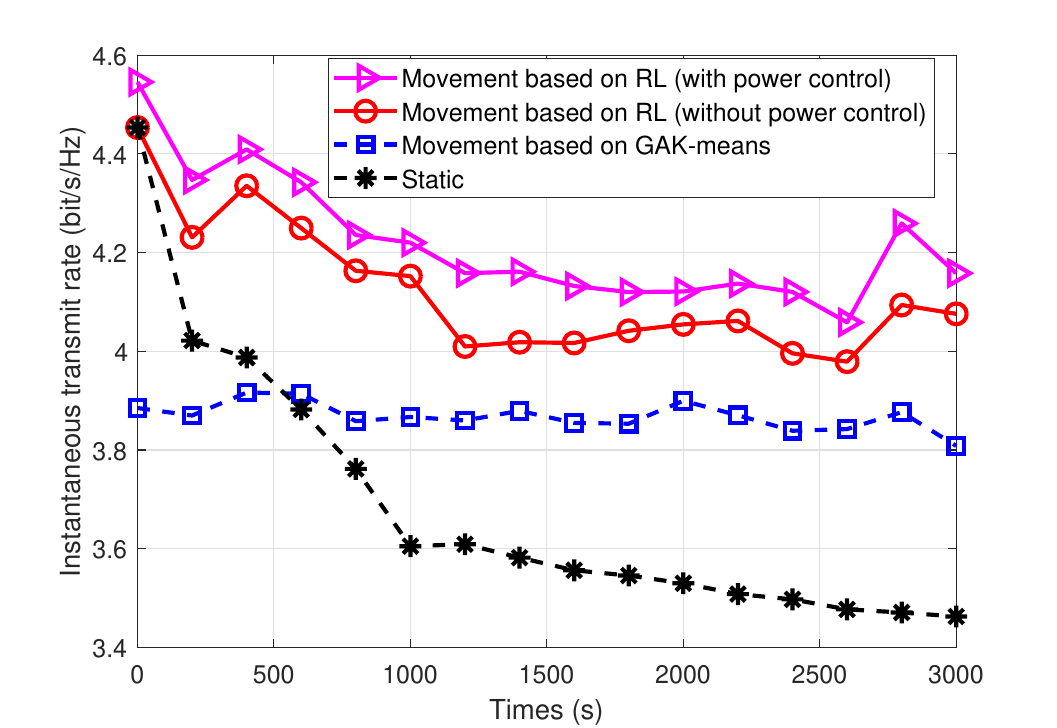}
 \caption{Comparison between movement and static scenario over throughput.}\label{transmitrate}
\end{figure}

\begin{figure*} \centering
\subfigure[Initial and final positions of the UAVs and the users.] { \label{fig:a}
\includegraphics[width=0.92\columnwidth]{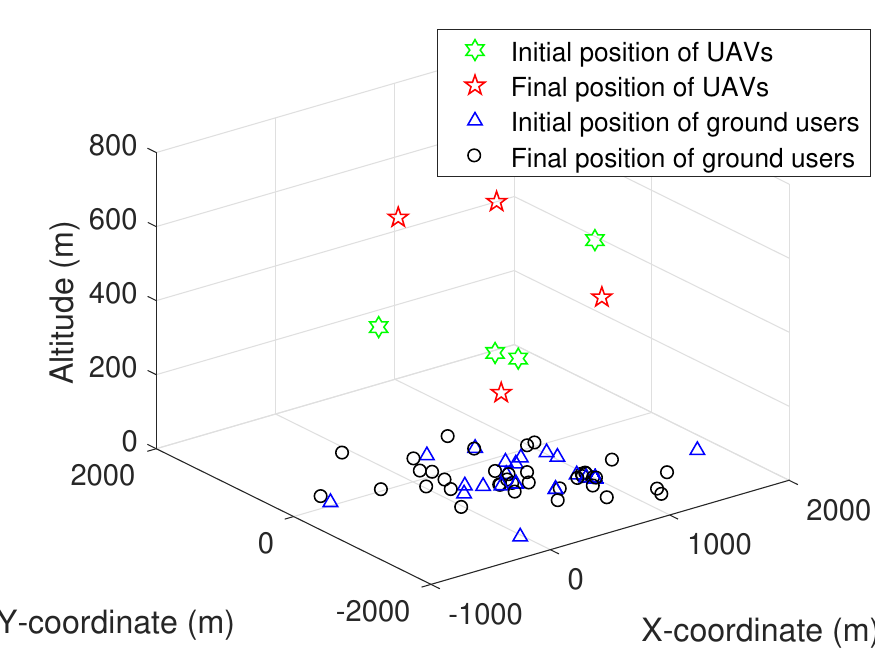}
}
\subfigure[Movement of one of the UAVs projected in two-dimensional.] { \label{fig:b}
\includegraphics[width=0.9\columnwidth]{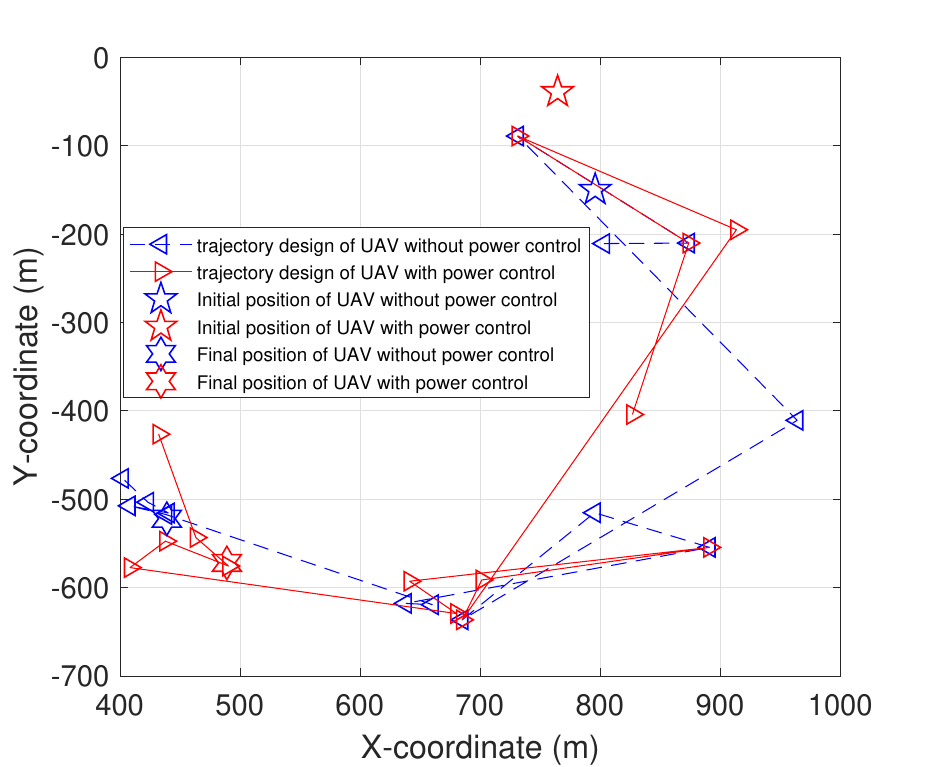}
}
\caption{Positions of the users and the UAVs as well as the trajectory design of UAVs both with and with out power control. }\label{path2}
\end{figure*}

In the multi-agent Q-learning model, the learning agent has to handle $N$ Q-functions, one for each agent in the model. These Q-functions are handled internally by the learning agent, assuming that it can observe other agents' actions and rewards. The learning agent updates $\left( {{Q^1},...,{Q^N}} \right)$, where each ${Q^n},n = 1,...,N$, is constructed of ${Q^n}\left( {s,{a^1},...,{a^N}} \right)$ for all ${s,{a^1},...,{a^N}}$. Assuming $\left| {{A^1}} \right| =  \cdots  = \left| {{A^N}} \right| = \left| A \right|$, where $\left| S \right|$ is the number of states, and $\left| {{A^n}} \right|$ is the size of agent $n$'s action space $A^n$. Then, the total number of entries in $Q^n$ is $\left| S \right| \cdot {\left| A \right|^n}$. Finally, the total storage space requirement is $N\left| S \right| \cdot {\left| A \right|^N}$. Therefore the space size of the model is increased linearly with the number of states, polynomially with the number of actions, but exponentially with the number of agents.

\section{Numerical Results}
Our simulation parameters are given in Table~\ref{table:parameters}. The initial locations of the UAVs are randomized. The maximum transmit power of each UAV is the same, and transmit power is uniformly allocated to users. On this basis, we analyze the instantaneous transmit rate of users, position prediction of the users, the 3D trajectory design and power control of the UAVs.

\begin{table}[htbp]
\caption{ Simulation parameters}
\centering
\begin{tabular}{|l||r||r|}
\hline
\centering
 Parameter & Description & Value\\
\hline
\centering
 $f_c$ & Carrier frequency & 2GHz\\
\hline
\centering
 $N_0$ & Noise power spectral & -170dBm/Hz\\
\hline
\centering
 $N_x$ & Size of neuron reservoir & 2000\\
\hline
\centering
 $N$ & Number of UAVs & 4\\
\hline
\centering
 $B$ & Bandwidth & 1MHz\\
\hline
\centering
 $b_1$,$b_2$ & Environmental parameters & 0.36,0.21 ~\cite{Mozaffari2017IEEE_J_WCOM}\\
\hline
\centering
 $\alpha $ & Path loss exponent & 2\\
 \hline
\centering
 ${\mu _{LoS}}$ & Additional path loss for LoS & 3dB ~\cite{Mozaffari2017IEEE_J_WCOM}\\
 \hline
\centering
 ${\mu _{NLoS}}$ & Additional path loss for NLoS & 23dB ~\cite{Mozaffari2017IEEE_J_WCOM}\\
 \hline
 \centering
 $\alpha_l $ & Learning rate & 0.01\\
 \hline
 \centering
 $\beta $ & Discount factor & 0.7\\
 \hline
\end{tabular}
\label{table:parameters}
\end{table}

\subsection{Predicted Users' Positions}

Fig.~\ref{esnplot} characterizes the prediction accuracy of a user's position parameterized by the reservoir size. It can be observed that increasing the reservoir size of the ESN algorithm leads to a reduced error between the real tracks and predicated tracks. Again, the larger the neuron reservoir size, the more precise the prediction becomes, but the probability of causing overfitting is also increased. This is due to the fact that the size of the ESN reservoir directly affects the ESN's memory requirement which in turn directly affects the number of user positions that the ESN algorithm is capable of recording. When the neuron reservoir size is 1000, a high accuracy is attained.

\begin{table}[thpb]
\caption{Performance Comparison Between ESN algorithm and benchmarks}
\centering
\begin{tabular}{|l||r||r||r||r|}
\hline
\centering
 Metric & HA & $\text{ESN}_{500}$ & $\text{ESN}_{1000}$ & LSTM\\
\hline
\centering
 MSE & 41.78 & 25.17 & 19.36 & 24.82 \\
\hline
\centering
Computing time & 116ms &161ms & 737ms & 2103ms\\
\hline
\end{tabular}
\label{table:performance}
\end{table}

Table III characterizes the performance of the proposed ESN model. The so-called historical average (HA) model and the long short term memory (LSTM) model are also used as our benchmarks. It can be observed that the ESN having a neuron reservoir size of 1000 attains a lower MSE than the HA model and the LSTM model, even though the complexity of the ESN model is far lower than that of the  LSTM model. Overall, the proposed ESN algorithm outperforms the benchmarks.

\subsection{Trajectory Design and Power Control of UAVs}

Fig.~\ref{convergence2} characterizes the throughput vs the number of training episodes. It can be observed that the UAVs are capable of carrying out their actions in an iterative manner and learn from their mistakes for improving the throughput. When three UAVs are employed, convergence is achieved after about 45000 episodes, whilst 30000 more training episodes are required for convergence when the number of UAV is four. Additionally, the learning rate of 0.80 used for the multi-agent Q-learning model outperforms that of 0.60 and 0.70 in terms of the throughput. Although the model relying on a learning rate of 0.90 converges faster than other models, this model is more likely to converge to a sub-optimal $Q^{*}$ value, which leads to a lower throughput.

Fig.~\ref{transmitrate} characterizes the throughput with the movement derived from multi-agent Q-learning. The throughput in the scenario that users remain static and the throughput with the movement derived by the GAK-means are also illustrated as benchmarks. It can be observed that the instantaneous transmit rate decreases as time elapses. This is because the users are roaming during each time slot. At the initial time slot, the users (namely the people who tweet) are flocking together around Oxford Street in London, but  after a few hundred seconds, some of the users move away from Oxford Street. In this case, the density of users is reduced, which affects the instantaneous sum of the transmit rate. It can also be observed that re-deploying UAVs based on the movement of users is an effective method of mitigating the downward trend compared the static scenario. Fig.~\ref{transmitrate} also illustrates that the movement of UAVs relying on power control is more capable of maintaining a high-quality service than the mobility scenario operating without power control. Additionally, it also demonstrates that the proposed multi-agent Q-learning based trajectory-acquiring and power control algorithm outperforms GAK-means algorithm also used as a benchmark.

\begin{figure}
\centering
     \includegraphics[width=3.2in]{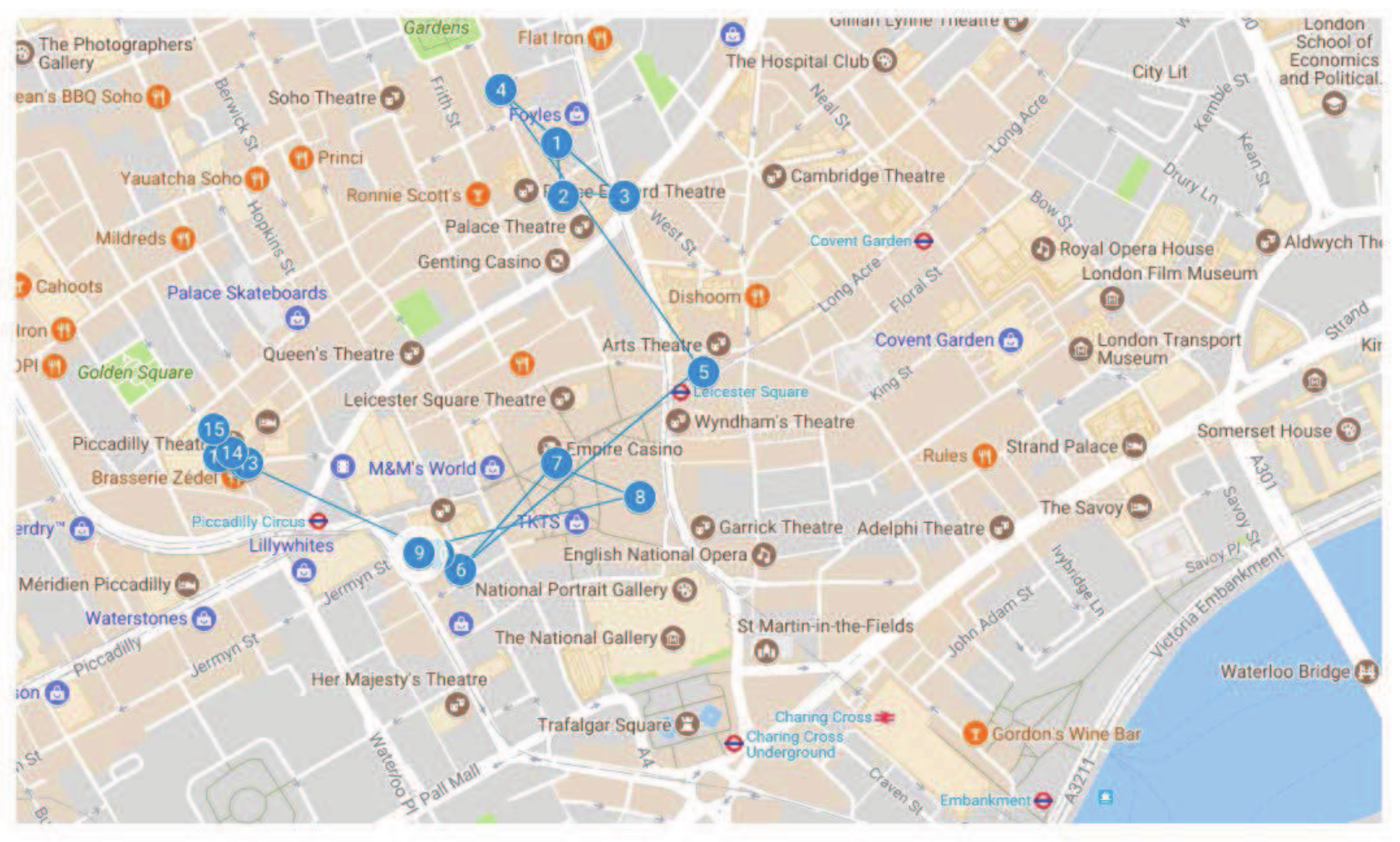}
     \caption{Trajectory design of one of the UAVs on Google Map.}\label{map2}
  \end{figure}

Fig.~\ref{path2} characterizes the designed 3D trajectory for one of the UAVs both in the scenario of moving with power control and in its counterpart operating without power control. Compared to only consider the trajectory design of UAVs, jointly consider both the trajectory design and the power control results in different trajectories for the UAVs. However, the main flying direction of the UAVs remains the same. This is because the interference is also considered in our model and power control of UAVs is capable of striking a tradeoff between increasing the received signal power and the interference power, which in turn increases the received SINR.

Fig.~\ref{map2} characterizes the  trajectory designed for one of the UAVs on Google map. The trajectory consists of 16 hevering points, where each UAV will stop for about 200 seconds. The number of the hovering points has to be appropriately chosen based on the specific requirements in the real scenario. Meanwhile, the trajectory of the UAVs may be designed in advance on the map with the aid of predicting the users' movements. In this case, the UAVs are capable of obeying a beneficial trajectory for maintaining a high quality of service without extra interaction from the ground control center.
\section{Conclusions}
The trajectory design and power control of multiple UAVs was jointly designed for maintaining a high quality of service. Three steps were provided for tackling the formulated problem. More particularly, firstly, multi-agent Q-learning based placement algorithm was proposed to deploy the UAVs at the initial time slot. Secondly, A real dataset was collected from Twitter for representing the users' position information and an ESN based prediction algorithm was proposed for predicting the future positions of the users. Thirdly, a multi-agent Q-learning based trajectory-acquisition and power-control algorithm was conceived for determining both the position and transmit power of the UAVs at each time slot. It was demonstrated that the proposed ESN algorithm was capable of predicting the movement of the users at a high accuracy. Additionally, re-deploying (trajectory design) and power control of the UAVs based on the movement of the users was an effective method of maintaining a high quality of downlink service.

\numberwithin{equation}{section}
\section*{Appendix~A: Proof of Lemma 1} \label{Appendix:A}
\renewcommand{\theequation}{A.\arabic{equation}}
\setcounter{equation}{0}
The rate requirement of each user is given by ${r_{{k_n}}}\left( t \right) \ge {r_0}$, then, we have\[{r_0} \le {B_{{k_n}}}\log 2\left( {1 + \frac{{{p_{{k_n}}}\left( t \right){g_{{k_n}}}\left( t \right)}}{{{I_{{k_n}}}\left( t \right) + {\sigma ^2}}}} \right).\tag{A.1}\]

Rewrite equation (A.2) as\[{p_{{k_n}}}\left( t \right) \ge \frac{{\left( {{I_{{k_n}}}\left( t \right) + {\sigma ^2}} \right)\left( {{2^{{{{r_0}} \mathord{\left/
 {\vphantom {{{r_0}} {{B_{{k_n}}}}}} \right.
 \kern-\nulldelimiterspace} {{B_{{k_n}}}}}}} - 1} \right)}}{{{g_{{k_n}}}\left( t \right)}}.\tag{A.2}\]

Then we have\[\begin{gathered}
  {P_{\max }} \ge \left| {{K_n}} \right|{K_0}d_{{k_n}}^a\left( t \right)\left( {{P_{{\text{LoS}}}}{\mu _{{\text{LoS}}}} + {P_{{\text{NLoS}}}}{\mu _{{\text{NLoS}}}}} \right)\hfill \\
{\kern 1pt}{\kern 1pt}{\kern 1pt}{\kern 1pt}{\kern 1pt}{\kern 1pt}{\kern 1pt}{\kern 1pt}{\kern 1pt}{\kern 1pt}{\kern 1pt}{\kern 1pt}{\kern 1pt}{\kern 1pt}{\kern 1pt}{\kern 1pt}{\kern 1pt}{\kern 1pt}{\kern 1pt}{\kern 1pt}{\kern 1pt}{\kern 1pt}{\kern 1pt}{\kern 1pt}{\kern 1pt}{\kern 1pt}{\kern 1pt}{\kern 1pt}{\kern 1pt}{\kern 1pt}{\kern 1pt}{\kern 1pt}{\kern 1pt}{\kern 1pt}\cdot \left( {{I_{{k_n}}}\left( t \right) + {\sigma ^2}} \right)\left( {{2^{\left| {{K_n}} \right|{{{r_0}} \mathord{\left/
 {\vphantom {{{r_0}} B}} \right.
 \kern-\nulldelimiterspace} B}}} - 1} \right) \hfill \\
\end{gathered} \tag{A.3}\]

It can be proved that $({P_{LoS}}{\mu _{LoS}} + {P_{NLoS}}{\mu _{NLoS}}) \le {\mu _{NLoS}}$, and the condition for equality is the probability of NLoS connection is 1. Following from the condition for equality, the maximize transmit rate of each UAV has to obey \[\begin{gathered}
 {P_{\max }}{\kern 1pt} {\kern 1pt} {\kern 1pt}  \ge \left| {{K_n}} \right|{\mu _{{\text{NLoS}}}}{\sigma ^2}{K_0}\left( {{2^{\left| {{K_n}} \right|{{{r_0}} \mathord{\left/
 {\vphantom {{{r_0}} B}} \right.
 \kern-\nulldelimiterspace} B}}} - 1} \right) \\
\cdot \max \left\{ {{h_1},{h_2}, \cdots {h_n}} \right\}
\end{gathered} \tag{A.4}\]

The proof is completed.

\numberwithin{equation}{section}
\section*{Appendix~B: Proof of Theorem 1} \label{Appendix:A}
\renewcommand{\theequation}{A.\arabic{equation}}
\setcounter{equation}{0}

Two steps are taken for proving the convergence of multi-agent Q-learning algorithm. Firstly, the convergence of single-agent model is proved. Secondly, we improve the results from the single-agent domain to the multi-agent domain.

The update rule of Q-learning algorithm is given by\[\begin{gathered}
  {Q_{t + 1}}({s_t},{a_t}) = \left( {1 - {\alpha _t}} \right){Q_t}\left( {{s_t},{a_t}} \right)  \hfill \\
{\kern 1pt}{\kern 1pt}{\kern 1pt}{\kern 1pt}{\kern 1pt}{\kern 1pt}{\kern 1pt}{\kern 1pt}{\kern 1pt}{\kern 1pt}{\kern 1pt}{\kern 1pt}{\kern 1pt}{\kern 1pt}{\kern 1pt}{\kern 1pt}{\kern 1pt}{\kern 1pt}{\kern 1pt}{\kern 1pt}{\kern 1pt}{\kern 1pt}{\kern 1pt}{\kern 1pt}{\kern 1pt}{\kern 1pt}{\kern 1pt}{\kern 1pt}{\kern 1pt}{\kern 1pt}{\kern 1pt}{\kern 1pt}{\kern 1pt}{\kern 1pt}{\kern 1pt}{\kern 1pt}{\kern 1pt}{\kern 1pt}{\kern 1pt}{\kern 1pt}{\kern 1pt}{\kern 1pt}{\kern 1pt}{\kern 1pt}{\kern 1pt}{\kern 1pt}{\kern 1pt}{\kern 1pt}{\kern 1pt}{\kern 1pt}{\kern 1pt}{\kern 1pt}+ {\alpha _t}\left[ {{r_t} + \beta \max {Q_t}\left( {{s_{t + 1}},{a_t}} \right)} \right] \hfill.\tag{B.1} \\
\end{gathered} \]

Subtracting the quantity ${Q^ * }({s_t},{a_t})$ from both side of the equation, we have \[\begin{gathered}
  {\Delta _t}({s_t},{a_t}) = {Q_t}({s_t},{a_t}) - {Q^ * }({s_t},{a_t}) \hfill \\
  {\kern 1pt} {\kern 1pt} {\kern 1pt} {\kern 1pt} {\kern 1pt} {\kern 1pt} {\kern 1pt} {\kern 1pt} {\kern 1pt} {\kern 1pt} {\kern 1pt} {\kern 1pt} {\kern 1pt} {\kern 1pt} {\kern 1pt} {\kern 1pt} {\kern 1pt} {\kern 1pt} {\kern 1pt} {\kern 1pt} {\kern 1pt} {\kern 1pt} {\kern 1pt} {\kern 1pt} {\kern 1pt} {\kern 1pt} {\kern 1pt} {\kern 1pt} {\kern 1pt} {\kern 1pt} {\kern 1pt} {\kern 1pt} {\kern 1pt} {\kern 1pt} {\kern 1pt} {\kern 1pt} {\kern 1pt} {\kern 1pt} {\kern 1pt} {\kern 1pt} {\kern 1pt} {\kern 1pt} {\kern 1pt}  = \left( {1 - {\alpha _t}} \right){\Delta _t}({s_t},{a_t})\hfill \\
{\kern 1pt} {\kern 1pt}{\kern 1pt} {\kern 1pt} {\kern 1pt}{\kern 1pt} {\kern 1pt} {\kern 1pt}{\kern 1pt} {\kern 1pt} {\kern 1pt}{\kern 1pt} {\kern 1pt} {\kern 1pt}{\kern 1pt} {\kern 1pt} {\kern 1pt}{\kern 1pt} {\kern 1pt} {\kern 1pt}{\kern 1pt} {\kern 1pt} {\kern 1pt}{\kern 1pt} {\kern 1pt} {\kern 1pt}{\kern 1pt} {\kern 1pt} {\kern 1pt}{\kern 1pt} {\kern 1pt} {\kern 1pt} {\kern 1pt}{\kern 1pt} {\kern 1pt} {\kern 1pt}{\kern 1pt} {\kern 1pt} {\kern 1pt}{\kern 1pt} {\kern 1pt} {\kern 1pt}{\kern 1pt} {\kern 1pt} {\kern 1pt}+ {\alpha _t}\left[ {{r_t} + \beta \max {Q_t}\left( {{s_{t + 1}},{a_{t + 1}}} \right) - {Q^ * }({s_t},{a_t})} \right] \hfill \\
\end{gathered}.\tag{B.2} \]

We write ${F_t}({s_t},{a_t}) = {r_t} + \beta \max {Q_t}\left( {{s_{t + 1}},{a_{t + 1}}} \right) - {Q^ * }({s_t},{a_t})$, then we have \[\begin{gathered}
  {\rm E}\left[ {{F_t}({s_t},{a_t})\left| {{\operatorname{F} _t}} \right.} \right] = \sum\limits_{{s_t} \in S} {{P_{{a_t}}}({s_t},{s_{t + 1}})}\hfill \\
 \times {\kern 1pt} \left[ {{r_t} + \beta \max {Q_t}\left( {{s_{t + 1}},{a_{t + 1}}} \right) - {Q^ * }({s_t},{a_t})} \right] \hfill \\
 = \left( {{\mathbf{{\rm H}}}{Q_t}} \right)({s_t},{a_t}) - {Q^ * }({s_t},{a_t}) \hfill \\
\end{gathered}.\tag{B.3} \]

Using the fact that ${\mathbf{{\rm H}}}{Q^ * } = {Q^ * }$, then, ${\rm E}\left[ {{F_t}({s_t},{a_t})\left| {{\operatorname{F} _t}} \right.} \right] = \left( {{\mathbf{{\rm H}}}{Q_t}} \right)({s_t},{a_t}) - \left( {{\mathbf{{\rm H}}}{Q^ * }} \right)({s_t},{a_t})$. It has been proved that ${\left\| {{\mathbf{{\rm H}}}{Q_1} - {\mathbf{{\rm H}}}{Q_2}} \right\|_\infty } \leq \beta \left\| {{Q_1} - {Q_2}} \right\|$~\cite{jaakkola1994convergence}. In this case, we have ${\left\| {{\rm E}\left[ {{F_t}({s_t},{a_t})\left| {{\operatorname{F} _t}} \right.} \right]} \right\|_\infty } \le \beta {\left\| {{Q_t} - {Q^ * }} \right\|_\infty } = \beta {\left\| {{\Delta _t}} \right\|_\infty }$.  Finally, \[\begin{gathered}
  {\mathbf{VAR}}\left[ {{F_t}({s_t},{a_t})\left| {{\operatorname{F} _t}} \right.} \right] \hfill \\
= {\rm E}\left[ {{{\left( {{r_t} + \beta \max {Q_t}\left( {{s_{t + 1}},{a_{t + 1}}} \right) - \left( {{\mathbf{{\rm H}}}{Q_t}} \right)({s_t},{a_t})} \right)}^2}} \right] \hfill \\
= {\mathbf{VAR}}\left[ {{r_t} + \beta \max {Q_t}\left( {{s_{t + 1}},{a_{t + 1}}} \right)\left| {{\operatorname{F} _t}} \right.} \right] \hfill \\
\end{gathered} .\tag{B.4}\]

Due to the fact that $r$ is bounded, clearly verifies ${\mathbf{VAR}}\left[ {{F_t}({s_t},{a_t})\left| {{\operatorname{F} _t}} \right.} \right] \le C\left( {1 + \left\| {{\Delta _t}} \right\|_{}^2} \right)$ for some constant $C$.

Then, as ${{\Delta _t}}$ converges to zero under the assumptions in~\cite{melo2001convergence}, the single model converges to the optimal Q-function as long as $0 \le {\alpha _t} \le 1,{\kern 1pt} {\kern 1pt} {\kern 1pt} {\kern 1pt} \sum\limits_t {{\alpha _t} = \infty {\kern 1pt} } $ and $\sum\limits_t {\alpha _t^2 < \infty {\kern 1pt} } $.

Then, we improve the results to multi-agent domain. We assume that, there is an initial card which contains an initial value $M{Q^{n * }}\left( {\left\langle {{s_n},0} \right\rangle ,{a_n}} \right)$ at the bottom of the multi deck. The $Q$ value for episode 0 in multi-agent algorithm has the same as an initial value, which is expressed as $M{Q^{n * }}\left( {\left\langle {{s_n},0} \right\rangle ,{a_n}} \right) = MQ_0^n\left( {{s_n},{a_n}} \right)$.

For episode $k$, an optimal value is equivalent to the $Q$ value, $M{Q^{n * }}\left( {\left\langle {{s_n},k} \right\rangle ,{a_n}} \right) = MQ_k^n\left( {{s_n},{a_n}} \right)$. Next, we consider a value function which selects optimal action by using an equilibrium strategy. At the $k$ level, an optimal value function is the same as the Q value function. \[\begin{gathered}
  {V^{n * }}\left( {\left\langle {{s_{n + 1}},k} \right\rangle } \right) = V_k^n\left( {{s_{n + 1}}} \right)\hfill \\
 = {\text{E}}Q_k^n\left[ {\prod\limits_{j = 1}^n {{\beta _j} \times \max MQ_k^n\left( {{s_{n + 1}},{a_{n + 1}}} \right)} } \right] \hfill \\
\end{gathered} .\tag{B.5}\]

One of the agents maintains the previous $Q$ value for episode $k+1$. Then, we have \[\begin{gathered}
  MQ_{k + 1}^n\left( {{s_n},{a_n}} \right) = MQ_k^n\left( {{s_n},{a_n}} \right)= MQ_k^{n * }\left( {\left\langle {{s_n},k} \right\rangle ,{a_n}} \right)\hfill \\
  {\kern 1pt} {\kern 1pt} {\kern 1pt} {\kern 1pt} {\kern 1pt} {\kern 1pt} {\kern 1pt} {\kern 1pt} {\kern 1pt} {\kern 1pt} {\kern 1pt} {\kern 1pt} {\kern 1pt} {\kern 1pt} {\kern 1pt} {\kern 1pt}  {\kern 1pt} {\kern 1pt} {\kern 1pt} {\kern 1pt} {\kern 1pt} {\kern 1pt} {\kern 1pt} {\kern 1pt} {\kern 1pt} {\kern 1pt} {\kern 1pt} {\kern 1pt} {\kern 1pt} {\kern 1pt} {\kern 1pt} {\kern 1pt} {\kern 1pt} {\kern 1pt} {\kern 1pt}  {\kern 1pt} {\kern 1pt} {\kern 1pt} {\kern 1pt} {\kern 1pt} {\kern 1pt} {\kern 1pt} {\kern 1pt} {\kern 1pt} {\kern 1pt} {\kern 1pt} {\kern 1pt} {\kern 1pt} {\kern 1pt} {\kern 1pt} {\kern 1pt} {\kern 1pt} {\kern 1pt} {\kern 1pt}{\kern 1pt} {\kern 1pt} {\kern 1pt} {\kern 1pt} {\kern 1pt} {\kern 1pt} {\kern 1pt} {\kern 1pt} {\kern 1pt} {\kern 1pt} {\kern 1pt} {\kern 1pt} {\kern 1pt} {\kern 1pt}{\kern 1pt}= MQ_k^{n * }\left( {\left\langle {{s_n},k + 1} \right\rangle ,{a_n}} \right) \hfill \\
\end{gathered} .\tag{B.6}\]

Otherwise, the $Q$ value holds the previous multi-agent $Q$ value with probability of $1 - \alpha _{k + 1}^n$ and takes two types of rewards with probability of $\alpha _{k + 1}^n$. Then we have \[\begin{gathered}
  MQ_{}^{n * }\left( {\left\langle {{s_n},k + 1} \right\rangle ,{a_n}} \right)\hfill \\
 = (1 - \alpha _{k + 1}^n)MQ_k^{n * }\left( {\left\langle {{s_n},k} \right\rangle ,{a_n}} \right) \hfill \\
+ \alpha _{k + 1}^n\left[ {r_{k + 1}^n + \beta \sum\limits_{{s_{n + 1}}} {P_{{s_n} \to {s_{n + 1}}}^n\left[ {{a_n}} \right]V_k^n({s_{n + 1}})} } \right] \hfill \\
= (1 - \alpha _{k + 1}^n)MQ_k^n\left( {{s_n},{a_n}} \right) \hfill \\
+ \alpha _{k + 1}^n\left[ {r_{k + 1}^n + \beta \sum\limits_{{s_{n + 1}}} {P_{{s_n} \to {s_{n + 1}}}^n\left[ {{a_n}} \right]V_k^n({s_{n + 1}})} } \right] \hfill \\
 = MQ_{k + 1}^n\left( {{s_n},{a_n}} \right) \hfill \\
\end{gathered} .\tag{B.7}\]

In this case, if $MQ_{k + 1}^n\left( {{s_n},{a_n}} \right)$ converge to an optimal valve $MQ_{}^{n * }\left( {\left\langle {{s_n},k + 1} \right\rangle ,{a_n}} \right)$, then a state equation of multi-agent Q-learning ${\mathbf{MQ}}_{k + 1}^{}$ converges to an optimal state equation ${\mathbf{M}}{{\mathbf{Q}}^ * }\left[ {k + 1} \right]$.

 The proof is completed.

\bibliographystyle{IEEEtran}
\bibliography{mybib}

\begin{thebibliography}{10}
\providecommand{\url}[1]{#1}
\csname url@samestyle\endcsname
\providecommand{\newblock}{\relax}
\providecommand{\bibinfo}[2]{#2}
\providecommand{\BIBentrySTDinterwordspacing}{\spaceskip=0pt\relax}
\providecommand{\BIBentryALTinterwordstretchfactor}{4}
\providecommand{\BIBentryALTinterwordspacing}{\spaceskip=\fontdimen2\font plus
\BIBentryALTinterwordstretchfactor\fontdimen3\font minus
  \fontdimen4\font\relax}
\providecommand{\BIBforeignlanguage}[2]{{%
\expandafter\ifx\csname l@#1\endcsname\relax
\typeout{** WARNING: IEEEtran.bst: No hyphenation pattern has been}%
\typeout{** loaded for the language `#1'. Using the pattern for}%
\typeout{** the default language instead.}%
\else
\language=\csname l@#1\endcsname
\fi
#2}}
\providecommand{\BIBdecl}{\relax}
\BIBdecl

\bibitem{xiao2019Globecom}
X.~Liu, Y.~Liu, and Y.~Chen, ``Machine learning aided trajectory design and
  power control of multi-{UAV},'' in \emph{IEEE Proc. of Global Commun. Conf.
  (GLOBECOM)}, 2019.

\bibitem{Zhou2018TVT}
Y.~Zhou, P.~L. Yeoh, H.~Chen, Y.~Li, R.~Schober, L.~Zhuo, and B.~Vucetic,
  ``Improving physical layer security via a {UAV} friendly jammer for unknown
  eavesdropper location,'' \emph{{IEEE} Trans. Veh. Technol.}, vol.~67, no.~11,
  pp. 11\,280--11\,284, Nov. 2018.

\bibitem{khawaja2017uav}
W.~Khawaja, O.~Ozdemir, and I.~Guvenc, ``{UAV} air-to-ground channel
  characterization for mmwave systems,'' \emph{arXiv preprint
  arXiv:1707.04621}, 2017.

\bibitem{cheng2018TVT}
F.~Cheng, S.~Zhang, Z.~Li, Y.~Chen, N.~Zhao, F.~R. Yu, and V.~C.~M. Leung,
  ``{UAV} trajectory optimization for data offloading at the edge of multiple
  cells,'' \emph{{IEEE} Trans. Veh. Technol.}, vol.~67, no.~7, pp. 6732--6736,
  Jul. 2018.

\bibitem{osseiran2014COM}
A.~Osseiran, F.~Boccardi, V.~Braun, Kusume \emph{et~al.}, ``Scenarios for {5G}
  mobile and wireless communications: the vision of the {METIS} project,''
  \emph{{IEEE} Commun. Mag.}, vol.~52, no.~5, pp. 26--35, May 2014.

\bibitem{zeng2016COM}
Y.~Zeng, R.~Zhang, and T.~J. Lim, ``Wireless communications with unmanned
  aerial vehicles: {Opportunities} and challenges,'' \emph{{IEEE} Commun.
  Mag.}, vol.~54, no.~5, pp. 36--42, May 2016.

\bibitem{wang2018joint}
Q.~Wang, Z.~Chen, and S.~Li, ``Joint power and trajectory design for
  physical-layer secrecy in the {UAV}-aided mobile relaying system,''
  \emph{arXiv preprint arXiv:1803.07874}, 2018.

\bibitem{yi2019unified}
W.~Yi, Y.~Liu, E.~Bodanese, A.~Nallanathan, and G.~K. Karagiannidis, ``A
  unified spatial framework for {UAV}-aided {MmWave} networks,'' \emph{arXiv
  preprint arXiv:1901.01432}, 2019.

\bibitem{kandeepan2014aerial}
S.~Kandeepan, K.~Gomez, L.~Reynaud, and T.~Rasheed, ``Aerial-terrestrial
  communications: terrestrial cooperation and energy-efficient transmissions to
  aerial base stations,'' \emph{{IEEE} Trans. Aerosp. Electron. Syst.},
  vol.~50, no.~4, pp. 2715--2735, Dec 2014.

\bibitem{Yang2018energy}
D.~Yang, Q.~Wu, Y.~Zeng, and R.~Zhang, ``Energy tradeoff in ground-to-{UAV}
  communication via trajectory design,'' \emph{{IEEE} Trans. Veh. Technol.},
  vol.~67, no.~7, pp. 6721--6726, Jul. 2018.

\bibitem{zhang2018cellular}
S.~Zhang, Y.~Zeng, and R.~Zhang, ``Cellular-enabled {UAV} communication: A
  connectivity-constrained trajectory optimization perspective,'' \emph{arXiv
  preprint arXiv:1805.07182}, 2018.

\bibitem{katikala2014google}
S.~Katikala, ``Google project loon,'' \emph{InSight: Rivier Academic Journal},
  vol.~10, no.~2, pp. 1--6, 2014.

\bibitem{Mozaffari2017IEEE_J_WCOM}
M.~Mozaffari, W.~Saad, M.~Bennis, and M.~Debbah, ``Wireless communication using
  unmanned aerial vehicles ({UAVs}): {Optimal} transport theory for hover time
  optimization,'' \emph{{IEEE} Trans. Wireless Commun.}, vol.~16, no.~12, pp.
  8052--8066, Dec 2017.

\bibitem{zhang2017Optimal}
X.~Zhang and L.~Duan, ``Optimal deployment of {UAV} networks for delivering
  emergency wireless coverage,'' \emph{arXiv preprint arXiv:1710.05616}, 2017.

\bibitem{mozaffari2015drone}
M.~Mozaffari, W.~Saad, M.~Bennis, and M.~Debbah, ``Drone small cells in the
  clouds: {Design}, deployment and performance analysis,'' in \emph{IEEE Proc.
  of Global Commun. Conf. (GLOBECOM)}, 2015, pp. 1--6.

\bibitem{van2016lte}
B.~Van~der Bergh, A.~Chiumento, and S.~Pollin, ``{LTE} in the sky: trading off
  propagation benefits with interference costs for aerial nodes,'' \emph{{IEEE}
  Commun. Mag.}, vol.~54, no.~5, pp. 44--50, May 2016.

\bibitem{zeng2017WCOM}
Y.~Zeng and R.~Zhang, ``Energy-efficient {UAV} communication with trajectory
  optimization,'' \emph{{IEEE} Trans. Wireless Commun.}, vol.~16, no.~6, pp.
  3747--3760, Mar 2017.

\bibitem{liu2018comp}
L.~Liu, S.~Zhang, and R.~Zhang, ``{CoMP} in the sky: {UAV} placement and
  movement optimization for multi-user communications,'' \emph{arXiv preprint
  arXiv:1802.10371}, 2018.

\bibitem{sun2017air}
R.~Sun and D.~W. Matolak, ``Air-ground channel characterization for unmanned
  aircraft systems part {II}: Hilly and mountainous settings.'' \emph{{IEEE}
  Trans. Veh. Technol.}, vol.~66, no.~3, pp. 1913--1925, 2017.

\bibitem{bing2017study}
L.~Bing, ``Study on modeling of communication channel of {UAV},''
  \emph{Procedia Computer Science}, vol. 107, pp. 550--557, 2017.

\bibitem{Zhang2018Synergy}
N.~Zhang, P.~Yang, J.~Ren, D.~Chen, L.~Yu, and X.~Shen, ``Synergy of big data
  and {5G} wireless networks: Opportunities, approaches, and challenges,''
  \emph{{IEEE} Wireless Commun.}, vol.~25, no.~1, pp. 12--18, Feb. 2018.

\bibitem{Yang2016Estimating}
B.~Yang, W.~Guo, B.~Chen, G.~Yang, and J.~Zhang, ``Estimating mobile traffic
  demand using {Twitter},'' \emph{{IEEE} Wireless Commun. Lett.}, vol.~5,
  no.~4, pp. 380--383, Aug. 2016.

\bibitem{Galindo2010Distributed}
A.~Galindo-Serrano and L.~Giupponi, ``Distributed {Q-Learning} for aggregated
  interference control in cognitive radio networks,'' \emph{{IEEE} Trans. Veh.
  Technol.}, vol.~59, no.~4, pp. 1823--1834, May 2010.

\bibitem{kosmerl2014ICC}
J.~Kosmerl and A.~Vilhar, ``Base stations placement optimization in wireless
  networks for emergency communications,'' in \emph{IEEE Proc. of International
  Commun. Conf. (ICC)}, 2014, pp. 200--205.

\bibitem{alhourani2014WCOML}
A.~Al-Hourani, S.~Kandeepan, and S.~Lardner, ``Optimal {LAP} altitude for
  maximum coverage,'' \emph{{IEEE} Wireless Commun. Lett.}, vol.~3, no.~6, pp.
  569--572, Jul 2014.

\bibitem{alzenad2017WCOML}
M.~Alzenad, A.~El-Keyi, F.~Lagum, and H.~Yanikomeroglu, ``3{D} placement of an
  unmanned aerial vehicle base station ({UAV-BS}) for energy-efficient maximal
  coverage,'' \emph{{IEEE} Wireless Commun. Lett.}, May 2017.

\bibitem{Sharma2017UAV}
P.~K. Sharma and D.~I. Kim, ``{UAV}-enabled downlink wireless system with
  non-orthogonal multiple access,'' pp. 1--6, Dec. 2017.

\bibitem{Sohail2018JA}
M.~F. Sohail, C.~Y. Leow, and S.~Won, ``Non-orthogonal multiple access for
  unmanned aerial vehicle assisted communication,'' \emph{{IEEE} Access},
  vol.~6, pp. 22\,716--22\,727, 2018.

\bibitem{hou2018multiple}
T.~Hou, Y.~Liu, Z.~Song, X.~Sun, and Y.~Chen, ``Multiple antenna aided {NOMA}
  in {UAV} networks: {A} stochastic geometry approach,'' \emph{arXiv preprint
  arXiv:1805.04985}, 2018.

\bibitem{Mozaffari2016WCOM}
M.~Mozaffari, W.~Saad, M.~Bennis, and M.~Debbah, ``Unmanned aerial vehicle with
  underlaid device-to-device communications: {Performance} and tradeoffs,''
  \emph{{IEEE} Trans. Wireless Commun.}, vol.~15, no.~6, pp. 3949--3963, Jun.
  2016.

\bibitem{mozaffari2016COML}
M.~Mozaffari, W.~Saad, and M.~Bennis, ``Efficient deployment of multiple
  unmanned aerial vehicles for optimal wireless coverage,'' \emph{{IEEE}
  Commun. Lett.}, vol.~20, no.~8, pp. 1647--1650, Jun 2016.

\bibitem{Lyu2016Cyclical}
J.~Lyu, Y.~Zeng, and R.~Zhang, ``Cyclical multiple access in {UAV-Aided}
  communications: {A} throughput-delay tradeoff,'' \emph{{IEEE} Wireless
  Commun. Lett.}, vol.~5, no.~6, pp. 600--603, Dec 2016.

\bibitem{Wu2018trajectory}
Q.~Wu, Y.~Zeng, and R.~Zhang, ``Joint trajectory and communication design for
  multi-{UAV} enabled wireless networks,'' \emph{{IEEE} Trans. Wireless
  Commun.}, vol.~17, no.~3, pp. 2109--2121, Mar. 2018.

\bibitem{wu2018common}
Q.~Wu and R.~Zhang, ``Common throughput maximization in {UAV}-enabled {OFDMA}
  systems with delay consideration,'' \emph{available online: arxiv.
  org/abs/1801.00444}, 2018.

\bibitem{zhang2017cellular}
S.~Zhang, Y.~Zeng, and R.~Zhang, ``Cellular-enabled {UAV} communication:
  {Trajectory} optimization under connectivity constraint,'' \emph{arXiv
  preprint arXiv:1710.11619}, 2017.

\bibitem{he2018joint}
H.~He, S.~Zhang, Y.~Zeng, and R.~Zhang, ``Joint altitude and beamwidth
  optimization for {UAV}-enabled multiuser communications,'' \emph{{IEEE}
  Commun. Lett.}, vol.~22, no.~2, pp. 344--347, 2018.

\bibitem{Liu2016Charging}
Q.~Liu, J.~Wu, P.~Xia, S.~Zhao, W.~Chen, Y.~Yang, and L.~Hanzo, ``Charging
  unplugged: Will distributed laser charging for mobile wireless power transfer
  work?'' \emph{{IEEE} Veh. Technol. Mag.}, vol.~11, no.~4, pp. 36--45, Dec.
  2016.

\bibitem{liu2018globalcom}
X.~Liu, Y.~Liu, and Y.~Chen, ``Deployment and movement for multiple aerial base
  stations by reinforcement learning,'' in \emph{IEEE Proc. of Global Commun.
  Conf. (GLOBECOM)}, Dec. 2018, pp. 1--6.

\bibitem{Ren2017Acess}
J.~Ren, G.~Zhang, and D.~Li, ``Multicast capacity for {VANETs} with directional
  antenna and delay constraint under random walk mobility model,'' \emph{{IEEE}
  Access}, vol.~5, pp. 3958--3970, 2017.

\bibitem{mingzhe2017JSAC}
M.~Chen, M.~Mozaffari, W.~Saad, C.~Yin, M.~Debbah, and C.~S. Hong, ``Caching in
  the sky: {P}roactive deployment of cache-enabled unmanned aerial vehicles for
  optimized quality-of-experience,'' \emph{{IEEE} J. Sel. Areas Commun.},
  vol.~35, no.~5, pp. 1046--1061, May 2017.

\bibitem{mingzhe2017esn}
M.~Chen, W.~Saad, C.~Yin, and M.~Debbah, ``Echo state networks for proactive
  caching in cloud-based radio access networks with mobile users,''
  \emph{{IEEE} Trans. Wireless Commun.}, vol.~16, no.~6, pp. 3520--3535, Jun.
  2017.

\bibitem{matignon2006reward}
L.~Matignon, G.~J. Laurent, and N.~Le~Fort-Piat, ``Reward function and initial
  values: better choices for accelerated goal-directed reinforcement
  learning,'' in \emph{International Conference on Artificial Neural Networks},
  2006, pp. 840--849.

\bibitem{jaakkola1994convergence}
T.~Jaakkola, M.~I. Jordan, and S.~P. Singh, ``Convergence of stochastic
  iterative dynamic programming algorithms,'' in \emph{Advances in neural
  information processing systems}, 1994, pp. 703--710.

\bibitem{melo2001convergence}
F.~S. Melo, ``Convergence of {Q-learning}: {A} simple proof,'' \emph{Institute
  Of Systems and Robotics, Tech. Rep}, pp. 1--4, 2001.

\end{thebibliography}

\end{document}